%% file: main.tex
\documentclass[10pt,journal,compsoc]{IEEEtran}
\usepackage{amsmath,amsfonts}
\usepackage{algorithm}
\usepackage{array}
\usepackage{textcomp}
\usepackage{stfloats}
\usepackage{url}
\usepackage{verbatim}
\usepackage{graphicx}
\usepackage{cite}
\usepackage{anyfontsize}
\usepackage{glossaries}
\usepackage{lscape}
\usepackage{rotating}
\usepackage{authblk}
\usepackage{amssymb}
\usepackage{multirow}
\usepackage{booktabs}
\input{macro}

\usepackage[left=0.5in, top=0.4in, right=0.3in, bottom=0.4in]{geometry}

\begin{document}

\title{Safeguarding Large Language Models: A Survey}

\author{Yi Dong$^{1*}$, Ronghui Mu$^{1*}$, Yanghao Zhang$^{1}$, Siqi Sun$^{1}$, Tianle Zhang$^{1}$, Changshun Wu$^{2}$, Gaojie Jin$^{1}$, Yi Qi$^{1}$, Jinwei Hu$^{1}$, Jie Meng$^{3}$, Saddek Bensalem$^{2,4}$, Xiaowei Huang$^{1}$
\thanks{* Equal contribution;\\
1. University of Liverpool, UK;\\
2. Université Grenoble Alpes, France;\\
3. Loughborough University, UK;\\
4. CSX-AI, France\\}
\thanks{Correspondence to: Xiaowei Huang <xiaowei.huang@liverpool.ac.uk>}
}
\markboth{IEEE TRANSACTIONS ON PATTERN ANALYSIS AND MACHINE INTELLIGENCE, VOL. XX, NO. X, MAY 2024}%
{Shell \MakeLowercase{\textit{et al.}}: A Sample Article Using IEEEtran.cls for IEEE Journals}

\IEEEtitleabstractindextext{
\begin{abstract}

In the burgeoning field of Large Language Models (LLMs), developing a robust safety mechanism, colloquially known as ``safeguards'' or ``guardrails'', has become imperative to ensure the ethical use of LLMs within prescribed boundaries. This article provides a systematic literature review on the current status of this critical mechanism. It discusses its major challenges and how it can be enhanced into a comprehensive mechanism dealing with ethical issues in various contexts. 
First, the paper elucidates the current landscape of safeguarding mechanisms that major LLM service providers and the open-source community employ. This is followed by the techniques to evaluate, analyze, and enhance some (un)desirable properties that a guardrail might want to enforce, such as hallucinations, fairness, privacy, and so on. Based on them, we review techniques to circumvent these controls (i.e., attacks), to defend the attacks, and to reinforce the guardrails. While the techniques mentioned above represent the current status and the active research trends, we also discuss several challenges that cannot be easily dealt with by the methods and present our vision on how to implement a comprehensive guardrail through the full consideration of multi-disciplinary approach, neural-symbolic method, and systems development lifecycle. 

\end{abstract}




}

\maketitle
\IEEEdisplaynontitleabstractindextext

%
\IEEEpeerreviewmaketitle

\IEEEraisesectionheading{\section{Introduction}\label{sec:1}}






In recent years, generative artificial intelligence (GenAI) has significantly accelerated humanity's stride into the era of intelligence. Technologies such as ChatGPT and Sora \cite{openai2023gpt4} have become a pivotal force driving the transformation of a new generation of industries.
However, the rapid deployment and integration of LLMs have raised significant concerns regarding their risks, including, but not limited to, ethical use, data biases, privacy, and robustness \cite{huangxiaowei2023survey}. In societal contexts, concerns also include the potential misuse by malicious actors through activities such as spreading misinformation or aiding criminal activities \cite{kang2023exploiting}.  
In the scientific context, LLMs can be used professionally, with dedicated ethical considerations and risks in scientific research \cite{birhane2023Science}.

To address these issues, model developers have implemented various safety protocols intended to confine the behaviors of these models to a more secure range of functions. 
The complexity of LLMs, characterized by intricate networks and numerous parameters, and the closed-source nature (such as ChatGPT) present substantial hurdles. These complexities require different strategies compared to the pre-LLM era,  which focuses on \emph{white-box techniques}, enhancing models by various regularizations and architecture adaptations during training. Therefore, in parallel to the reinforcement learning from human feedback (RLHF) and other training skills such as in-context training, the community moves towards employing \emph{black-box, post-hoc strategies}, notably \textbf{guardrails} \cite{welbl2021challenges,gehman2020realtoxicityprompts}, which monitors and filters the inputs and outputs of trained LLMs. A guardrail is an algorithm that takes as input a set of objects (e.g., the input and the output of LLMs) and determines if and how some enforcement actions can be taken to reduce the risks embedded in the objects. If the input to LLMs relates to child exploitation, the guardrail may stop the input or adapt the output to become harmless \cite{perez2022red}. In other words, guardrails are used to identify the potential misuse in the query stage and to prevent the model from providing an answer that should not be given. 

The difficulty in constructing guardrails often lies in establishing their requirements. AI regulations can be different across different countries, and in the context of a company, data privacy can be less severe than in the public domain. Nevertheless, a guardrail of LLMs may include \textbf{requirements} from one or more categories: Hallucination, fairness, privacy, robustness, toxicity, legality, out-of-distribution, uncertainty, etc.
In this paper, we do not include the typical requirement, i.e., accuracy, as they are benchmarks of the LLMs and arguably not the responsibilities of the guardrails. That said, there might not be a clear cut on the responsibilities (notably, robustness) between LLMs and the guardrails, and the two models shall collaborate to achieve a joint set of objectives. Nevertheless, for concrete applications, the requirements need to be precisely defined, and their corresponding metrics, and a \emph{multi-disciplinary} approach is called for. 
Mitigating a given requirement (such as hallucinations, toxicity, fairness, biases, etc) is already non-trivial, as discussed in Section~\ref{sec:5}. Working with multiple requirements worsens it, especially when some can be \emph{conflicting}. Such complexity requires a sophisticated solution design method to manage. 
In terms of the design of guardrails, while there might not be ``one method that rules them all'', a plausible design of the guardrail is \emph{neural-symbolic}, with learning agents and symbolic agents collaborating in processing both the inputs and the outputs of LLMs.  Multiple types of neural-symbolic agents   \cite{10.5555/3491440.3492119}. However, the existing guardrail solutions such as Llama Guard \cite{inan2023llama}, Nvidia NeMo \cite{rebedea2023nemo}, and Guardrails AI \cite{GuardrailsAI2023} use the simplest, loosely coupled ones. Given the complexity of the guardrails, it will be interesting to investigate other, more deeply coupled, neural-symbolic solution designs. 

Like safety-critical software, a \emph{systematic process} to cover the 
development cycle (ranging from specification to design, implementation, integration, verification, validation, and production release) is required to carefully build the guardrails, as indicated in industrial standards such as ISO-26262 and DO-178B/C. This survey starts with some background introduction at Section~\ref{sec:2}. The \textbf{goal} is to (1) Understand the existing guardrail frameworks that are being used to control model outputs in LLM services, as well as the techniques to evauate, analyze, and enhance guardrails against specific desirable properties (Section~\ref{sec:3}); (2) Understand the techniques that are being used to overcome these guardrails, as well as to defend the attacks and to reinforce the guardrails (Section~\ref{sec:4}); and then discuss how to achieve a complete guardrail solution, including several issues regarding the systematic design of a guardrail for a specific application context (Section~\ref{sec:5}). 



\section{Background for Large Language Models}\label{sec:2}




Large Language Models (LLMs), primarily based on the Transformer architecture \cite{NIPS2017_3f5ee243}, are composed of deep neural networks with several transformer blocks. Each block integrates a self-attention layer and a feedforward layer connected via residual links. This specific self-attention mechanism enables the model to concentrate on neighboring tokens while analyzing a specific token. Originally, the transformer architecture was exclusively developed for machine translation purposes. Newly developed language models that utilize transformer architecture can be fine-tuned, thereby removing the need for architectures tailored to specific tasks \cite{devlin-etal-2019-bert}. Typically, their networks encompass hundreds of billions (or more) of parameters and are trained on vast corpora of textual data. Examples include \textbf{ChatGPT-3} \cite{10.5555/3495724.3495883}, \textbf{ChatGPT-4} \cite{openai2023gpt4}, \textbf{LLaMA} \cite{touvron2023llama}, and \textbf{PaLM} \cite{anil2023palm}. 

LLMs are employed in a variety of complex tasks, such as conversational AI \cite{wei2023leveraging}, translation \cite{lyu2023new}, and story generation \cite{simon2022tattletale}. Current LLMs utilize architectures and training objectives similar to those in smaller language models, such as the Transformer architecture and tasks centered around language modeling. However, LLMs distinguish themselves by significantly scaling up in aspects like model dimensions, data volume, the breadth of their application scope, and computation cost. Building an offline model comprises three main stages \cite{huangxiaowei2023survey}: pre-training, adaptation tuning, and utilization improvement. Generally, the pre-training phase parallels conventional machine learning training, involving data collection, choosing an architecture, and undergoing training. The adaptation tuning includes instruction tuning \cite{lou2023prompt} and alignment tuning \cite{ouyang2022training} to enable learning from task-specific instructions and adhere to human values. Finally, the utilization improvements can enhance user interactions, including in-context learning \cite{10.5555/3495724.3495883} and chain-of-thought learning \cite{wei2022chain}.

After training an LLM, its performance against set expectations is crucial. 
This evaluation typically encompasses three dimensions: assessing essential performance, conducting safety analysis to understand potential consequences in practical applications, and utilizing publicly available benchmark datasets. 
The primary performance review focuses on essential capabilities like language generation and complex reasoning. 
Safety analysis delves into the LLM's alignment with human values, interactions with external environments, and integration into broader applications such as search engines. 
Additionally, benchmark datasets and accessible tools support this comprehensive evaluation. 
The outcome of this assessment determines whether the LLM meets pre-defined criteria and is ready for deployment. 
If it falls short, the process reverts to one of the earlier training stages to address identified shortcomings.
At the deployment stage, LLM could be used on a web platform for direct user interaction, like ChatGPT, or integrated into a search engine, like the new Bing. 
Regardless of the application, it is standard practice to implement guardrails in interactions between LLMs and users to ensure adherence to AI regulations.


\section{Techniques on Design and Implementation of Guardrails} \label{sec:3}

This section presents 
several existing guardrail techniques have been proposed by the LLM service provider or the open-source community. Then, we review the methods to evaluate, analyze, and enhance the LLMs according to the desirable properties one may expect an LLM to have. A comparison checkbox table among different platforms and properties is shown in Table.~\ref{compare_results}.


\begin{table}[htbp]
\caption{Abilities among different Guardrails.}\label{compare_results}
\resizebox{\columnwidth}{!}{\begin{tabular}{lcccccc}
\toprule
                        & Llama Guard  & Nvidia NeMo  & Guardrails AI & TruLens & Guidance AI & LMQL\\ 
\midrule
Hallucination    & \checkmark   &\checkmark&\checkmark   &\checkmark&\checkmark&\checkmark  \\
Fairness         & \checkmark   &-&\checkmark   &\checkmark&-&-  \\
Privacy          &-&\checkmark&- &-&-&- \\
Robustness       &-&-&- &-&-&- \\
Toxicity         &\checkmark&\checkmark&\checkmark  &\checkmark&\checkmark&\checkmark  \\ 
Legality         &\checkmark&-&- &-&-&-\\
Out-of-Distribution &-&-&\checkmark &-&-&-\\
Uncertainty      &-&\checkmark&\checkmark &\checkmark&-&- \\
\bottomrule
\end{tabular}}
\end{table}

\subsection{Guardrail Frameworks and Supporting Software Packages} \label{sec:guard}

\begin{figure}[htbp]
    \centering
    \includegraphics[width=\linewidth]{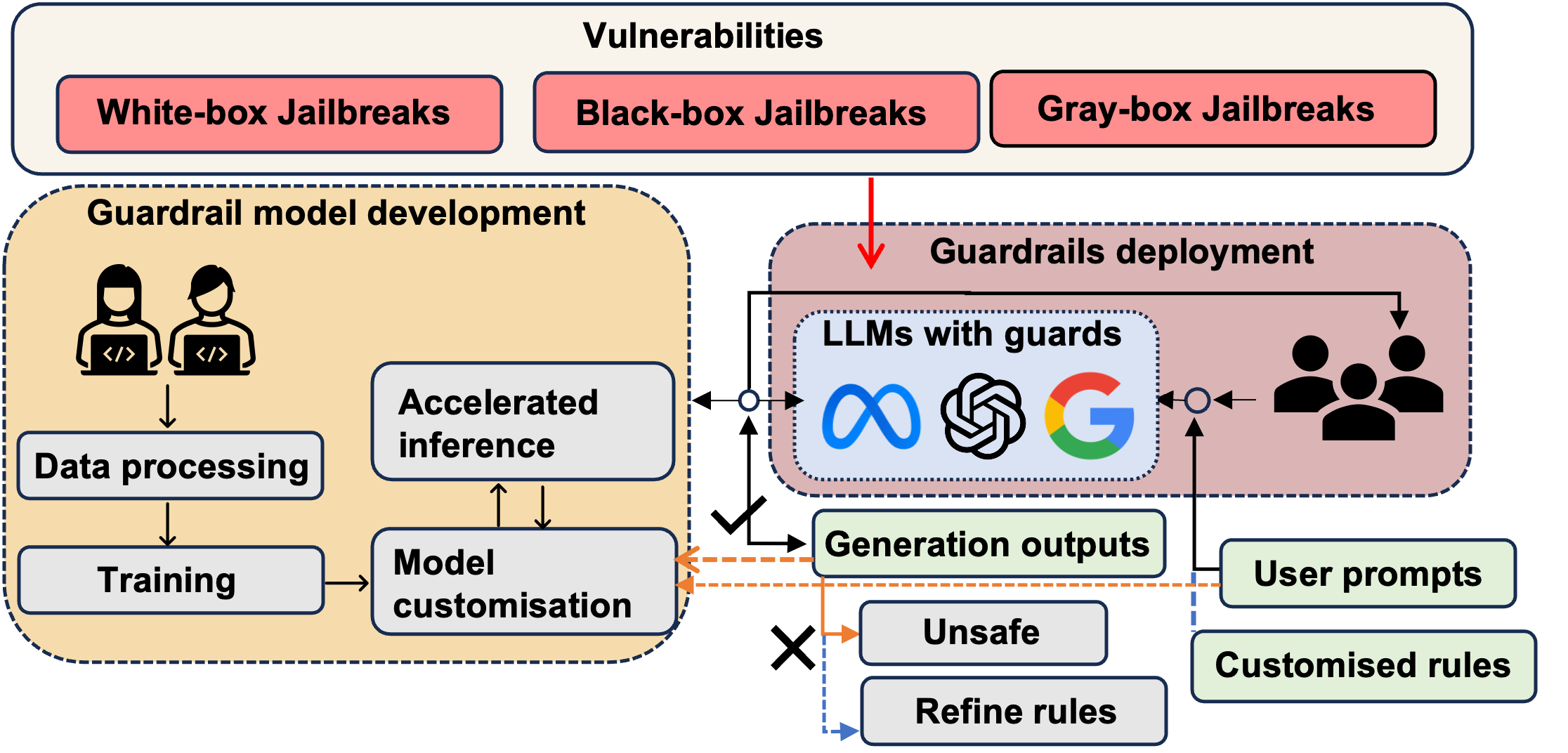}
    \caption{Guardrails Lifecycle and Vulnerabilities}
    \label{fig:guardrail_lifecycle}
\end{figure}

LLM guardrails constitute a suite of safety measures designed to oversee and regulate user interactions with LLM applications. 
These measures are programmable, rule-based systems positioned between users and foundational models. 
Their primary function is to ensure that the LLM model adheres to an organization's established principles and operates within a prescribed ethical and operational framework.
Guardrails are applied during the interaction stage between users and deployed LLMs, the last step in the LLM lifecycle. Fig. \ref{fig:guardrail_lifecycle} illustrates the lifecycle and potential vulnerabilities of the general guardrail mechanism. Developers complete the development of guardrails through data processing, guardrail model training, and model customization or fine-tuning (e.g., Llama Guard and NeMo Guardrails), as shown in the yellow area of Fig. \ref{fig:guardrail_lifecycle}. These guardrails are then deployed in LLMs to facilitate interaction with users. Typically, users predefine the content that needs protection, also called custom rules. Subsequently, users interact with LLMs through prompts and await the generated responses. The guardrails evaluate the output against the predefined rules to determine its compliance. If the content is deemed unsafe, guardrails may block it outright or issue a preset warning to the user. Conversely, if the output aligns with the criteria, it is displayed directly to the user, as indicated in the orange area of Fig. \ref{fig:guardrail_lifecycle}. Notably, some existing attack methods allow unsafe content to bypass guardrail protection, as highlighted in the red box of Fig. \ref{fig:guardrail_lifecycle}; for a detailed discussion of these attack methods, refer to Section \ref{sec:4}.

\subsubsection{Llama Guard} 
\begin{figure}[htbp]
    \centering
    \includegraphics [width=\columnwidth]{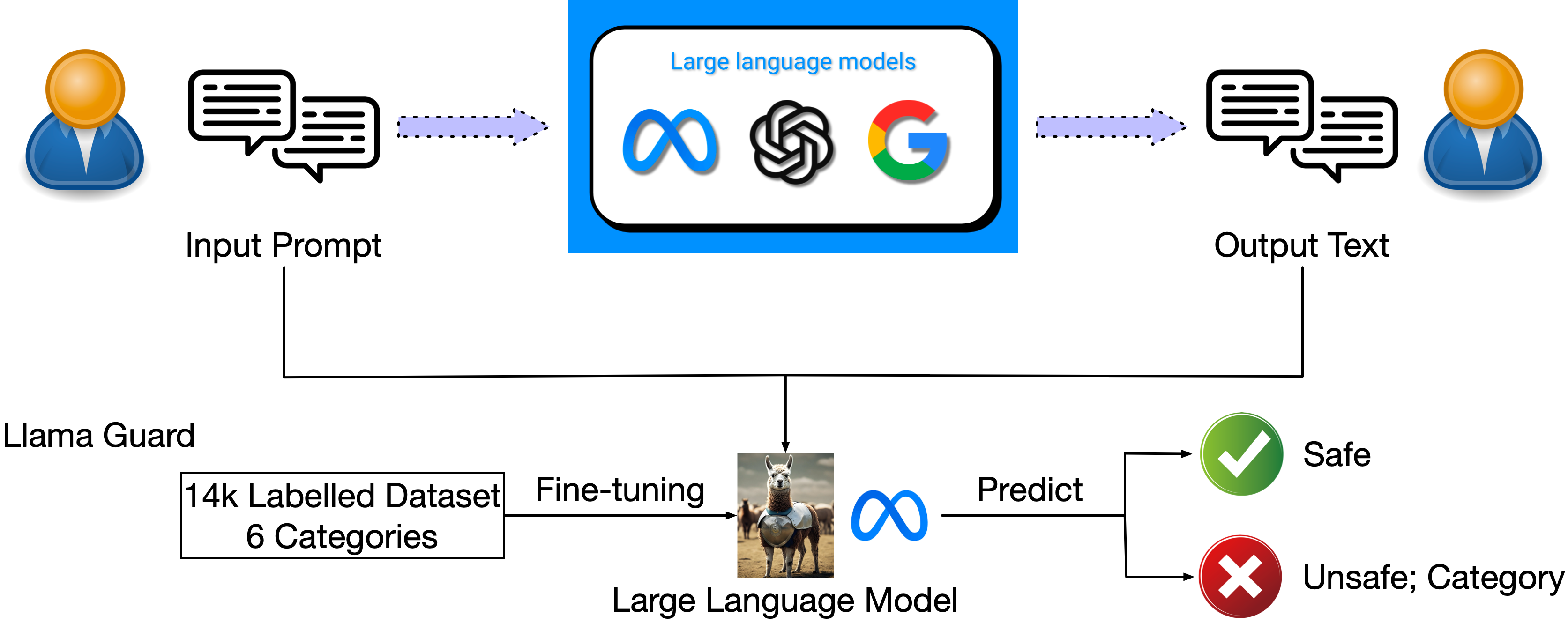}
    \caption{Llama Guard Guardrail Workflow}
    \label{fig:lg_llm}
\end{figure}

Llama Guard \cite{inan2023llama}, developed by Meta on the Llama2-7b architecture, focuses on enhancing human-AI conversation safety. 
It is a fine-tuned model that takes the input and output of the victim model as input and predicts their classification on a set of user-specified categories.
Figure \ref{fig:lg_llm} shows its workflow.
Due to the zero/few-shot  abilities of LLMs, Llama Guard can be adapted--by defining the user-specified categories 
--to different taxonomies and guidelines that meet requirements for applications and users. This is a Type 1 neural-symbolic system \cite{10.5555/3491440.3492119}, i.e., typical deep learning methods where the input and output of a learning agent are symbolic. It lacks guaranteed reliability since the classification results depend on the LLM's understanding of the categories and the model's predictive accuracy. 

\subsubsection{Nvidia Nemo} 
Nvidia NeMo, described in Rebedea's work 
\cite{rebedea2023nemo}, functions as an intermediary 
layer that enhances the control and safety of \gls{LLMs}. It employs Colang, an executable program language designed by 
\cite{Colang}, to establish constraints to guide LLMs within set dialogical boundaries. 
When the customer's input prompt comes, NeMo embeds the prompt as a vector and then uses \gls{KNN}
method to compare it with the stored vector-based user canonical forms, retrieving the embedding vectors that are `the most similar' to the embedded input prompt. The input embedding of Nemo differs from traditional approaches that utilize the initial layers. Instead, it employs embedding through similarity functions to capture the most relevant semantics. For example, Nemo uses the "sentence-transformers/all-MiniLM-L6-v2” model to compute embeddings, which is  used for the following KNN search (Annoy algorithm is employed for efficient nearest-neighbor search.) Therefore, ``input embedding” in the Nemo refers to mapping input sentences and paragraphs to a multi-dimensional dense vector space, facilitating the search for the most similar canonical forms/flows. After that, Nemo starts the flow execution to generate output from the canonical form. During the flow execution process, the LLMs generate a safe answer if the Colang program requests. 

LLMs will be invoked multiple times during the guardrail flow for various tasks. For example, in a conversation scenario, LLM is utilized in the following three phases:
(1) Generating user intent: the input of this LLM call includes two contexts: examples and potential user intents (top 5 intents from example code in NeMo Github repository). The output is a refined user intent (temperature is set as 0 to get the deterministic result). (2) Generating next step: In this phase, Nemo searches the most relevant similar flows and integrates these similar flows into an example, which is then fed into the LLM. The LLM call output is called ``bot intent.''.
 (3) Generating bot-message: The input for this call includes examples (the five most relevant bot intents) and relevant chunks (dictionary search), which are to be used as context.

\begin{figure}[htbp]
    \centering
    \includegraphics [width=\columnwidth]{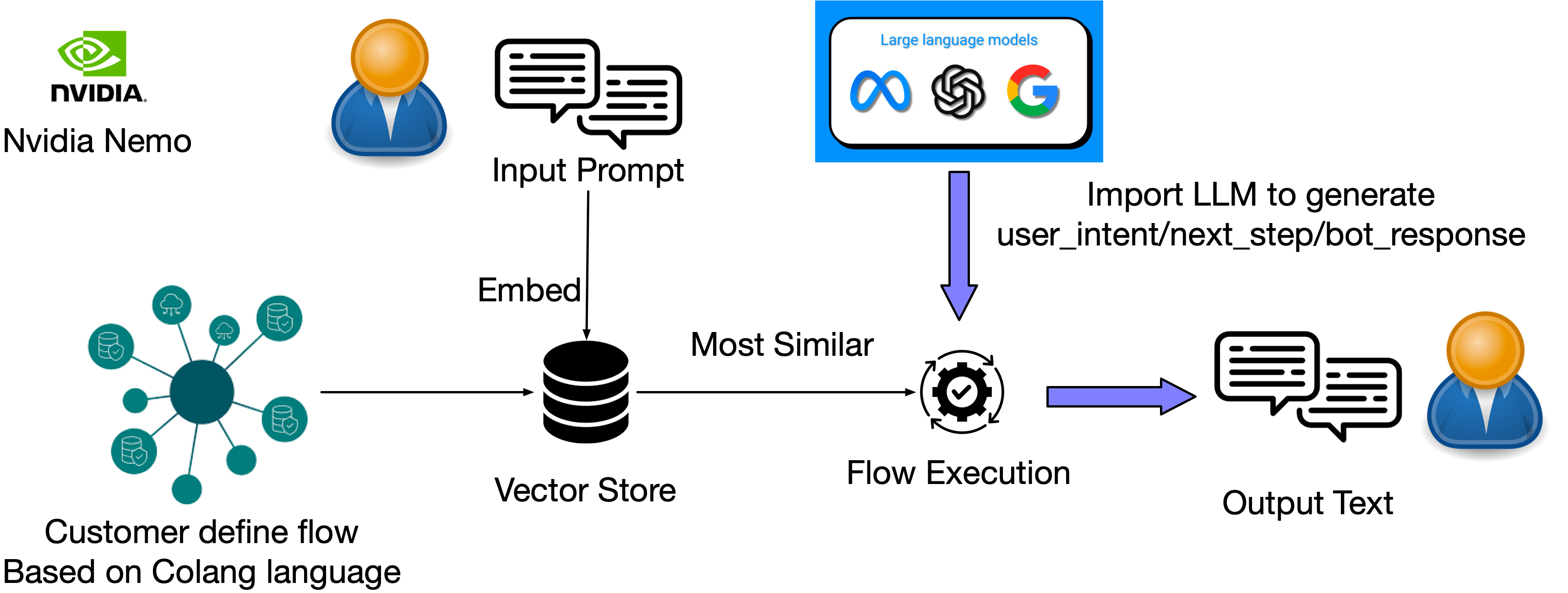}
    \caption{Nvidia NeMo Guardrails Workflow 
    }
    \label{fig:NN_llm}
\end{figure}

The process is presented in Figure \ref{fig:NN_llm}.
Building on the above customizable workflow, NeMo also includes a set of pre-implemented moderations dedicated to e.g.,  
fact-checking, hallucination prevention in responses, and content moderation. 
NeMo is also a Type-1 neural-symbolic system, with its effectiveness closely tied to the performance of the KNN method.

\subsubsection{Guardrails\ AI} 
Guardrails\ AI 
enables the user to add structure, type, and quality guarantees to the outputs of LLMs \cite{GuardrailsAI2023}. 
It operates in three steps: 1) defining the ``RAIL'' spec, 2) initializing the ``guard'' and 3) wrapping the LLMs.
In the first step, Guardrails AI defines a set of RAIL specifications, which are used to describe the return format limitations. This information must be written in a specific XML format, facilitating subsequent output checks, e.g., structure and types. The second step involves activating the defined spec as a guard. For applications that require categorized processing, such as toxicity checks, additional classifier models can be introduced to classify the input and output text. The third step is triggered when the guard detects an error. Here, the Guardrails AI can automatically generate a corrective prompt, pursuing the LLMs to regenerate the correct answer.  The output is then re-checked to ensure it meets the specified requirements. 
Currently, the methods based on Guardrails AI are only applicable for text-level checks and cannot be used in multimodal scenarios involving images or audio. Unlike the previous two methods, Guardrail AI is a Type-2 neural-symbolic system with a backbone symbolic algorithm supported by learning algorithms (in this case, those additional classifier models). 

\begin{figure}[htbp]
    \centering
    \includegraphics [width=\columnwidth]{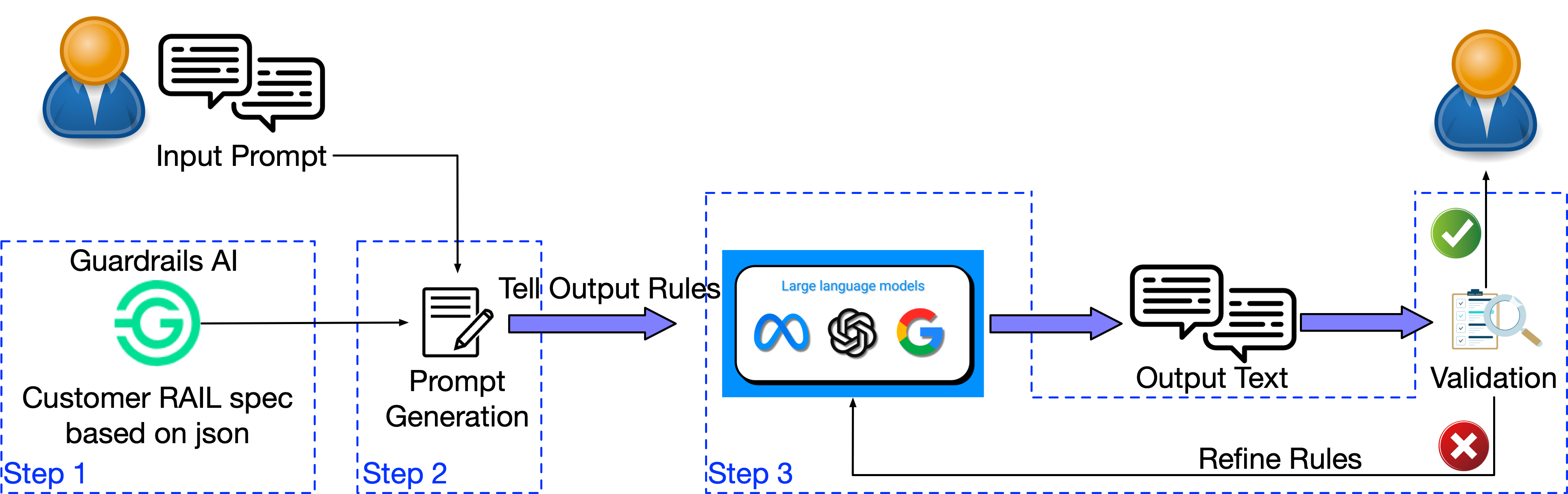}
    \caption{Guardrails AI Workflow}
    \label{fig:GA_llm}
\end{figure}

\subsubsection{TruLens} 
TruLens, developed by TruEra\footnote{https://truera.com/}, is an open-source toolkit for developing, evaluating, and monitoring LLMs. Central to its features is TruLens-Eval, which ensures quality by comparing outputs against prefined standards. The toolkit integrates LLMs, supporting logging records of inputs and outputs, and it leverages feedback functions that utilize auxiliary models, such as relevance models or sentiment classifiers, to perform evaluations on the RAG triad, including context relevance, answer relevance, and groundedness. Retrieval-Augmented Generation (RAG) is a technology that enhances the quality of language model outputs by supporting answer generation with retrieved relevant information. Within TruLens, RAG's role is to ensure the accuracy and relevance of model outputs by comparing them against predefined standards, thereby evaluating LLM apps. The services are invoked from various providers. For instance, when assessing groundedness related to how closely outputs align with the source material, Trulens-Eval can utilize providers like OpenAI API to call an LLM to find the relevant strings in a text or employ NLI models with hugging faces. The toolkit allows for the customization or pre-definition of feedback functions via Python, enabling evaluations to be specifically tailored to unique requirements. 
Additionally, TruLens incorporates embedding models to convert predefined information into numerical vectors, simplifying matching text with relevant data. TruLens also visualizes the LLM applications' rankings in a leaderboard according to their performance metrics, creating a dynamic environment that encourages developers to refine their models iteratively. As a guidance-oriented approach, TruLens is designed not to constrain LLM inputs and outputs but to provide a framework for continuous model refinement and evaluation, ensuring adherence to quality and relevance standards.

\begin{figure}[htbp]
    \centering
    \includegraphics [width=\columnwidth]{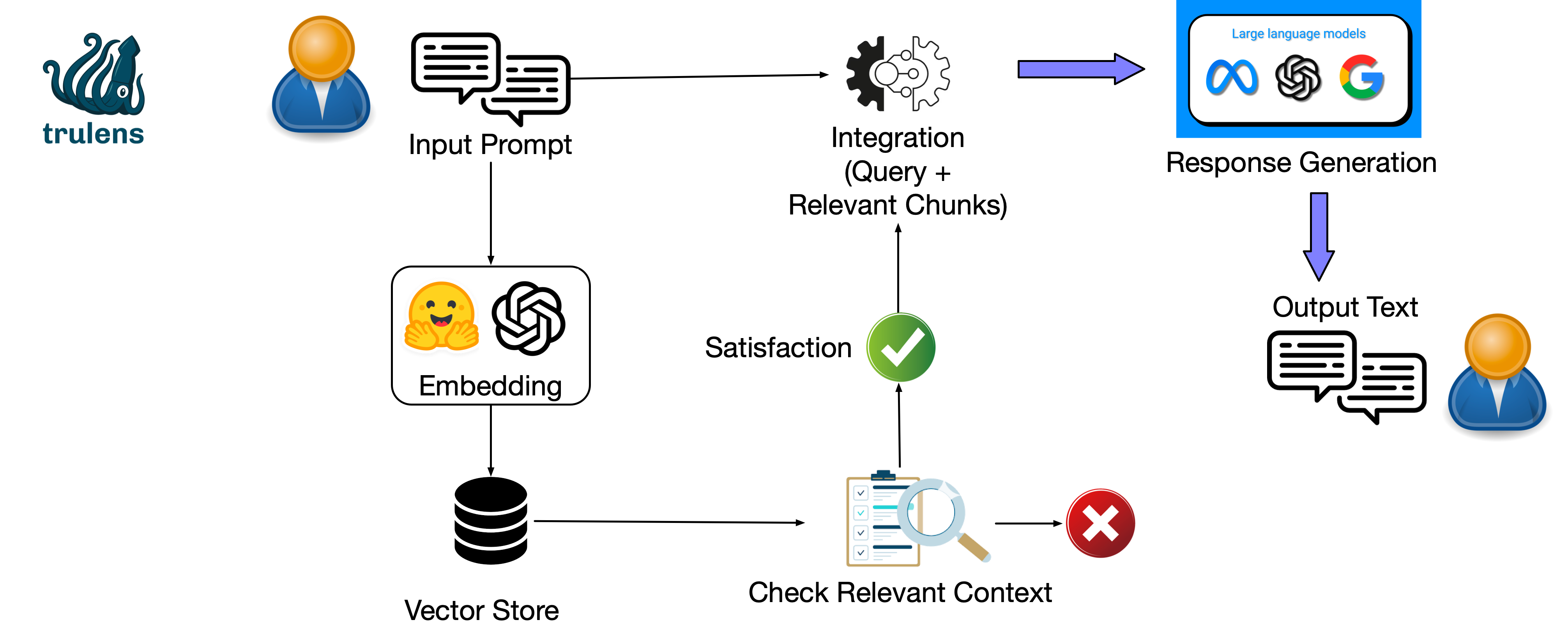}
    \caption{TruLens Workflow}
    \label{fig:TL_llm}
\end{figure}

\subsubsection{Guidance AI} 
\begin{figure}[htbp]
    \centering
    \includegraphics [width=\columnwidth]{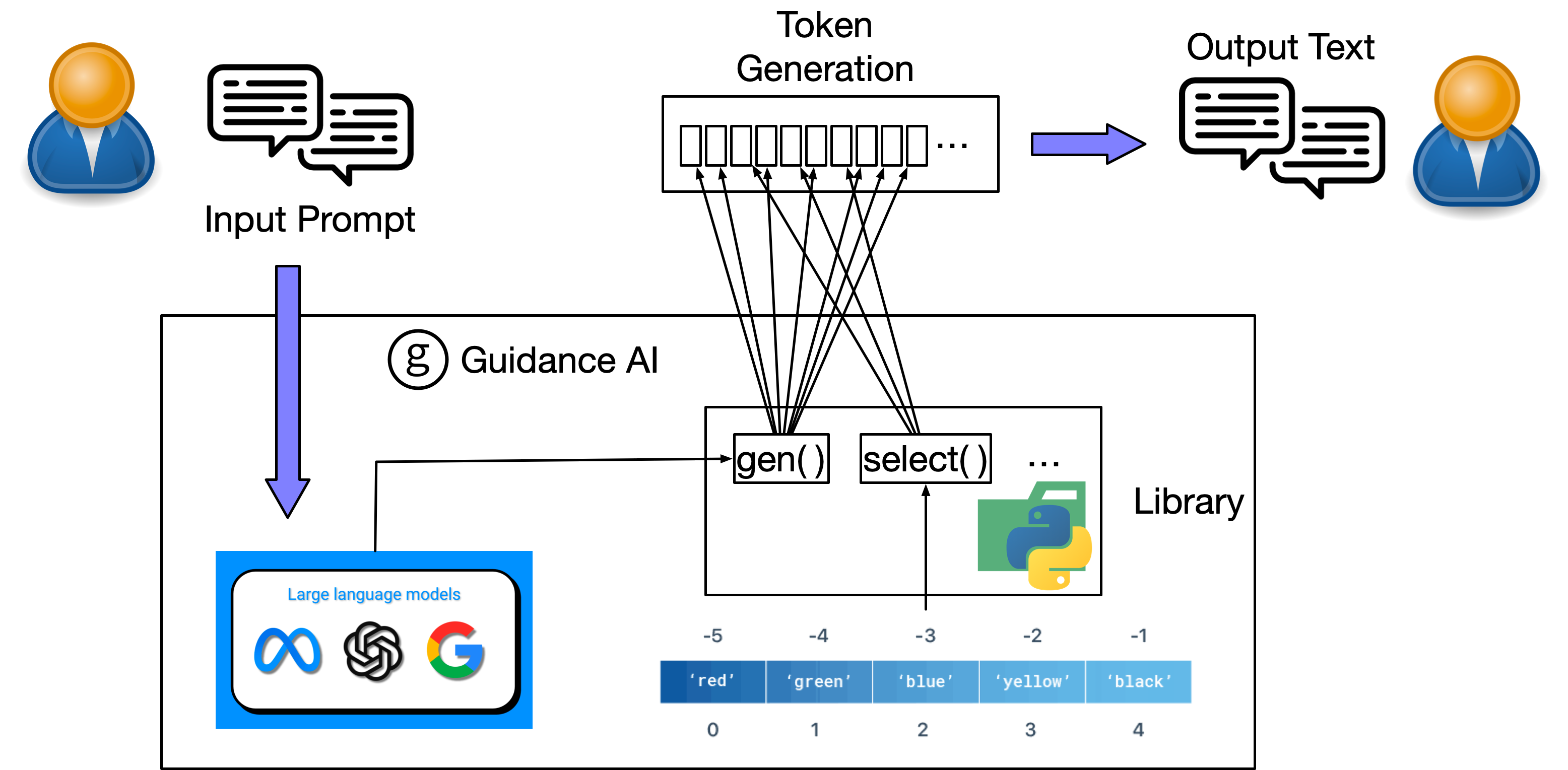}
    \caption{Guidance AI Workflow}
    \label{fig:GAI}
\end{figure}

Guidance AI\footnote{https://guidance.readthedocs.io/en/latest/}, a programming paradigm, offers superior control and efficiency than conventional prompting and chaining.  It allows users to constrain generation (e.g., with regex and CFGs) and interleave control (conditional, loops) and generation seamlessly.
This guardrail tool integrates text generation, prompts, and logic control within a single, continuous flow in a Python environment, thereby refining the text processing approach in LLMs. This unified method allows more effective LLM control than traditional prompts or thought language chains. Its features include simple and intuitive syntax built on the Handlebars template language, assuring the variable insertion in any prompts. The Guidance program has a well-defined linear execution order directly corresponding to the token sequence processed by the language model. The illustration graph of Guidance AI working flow is demonstrated in Figure~\ref{fig:GAI}.

At any timesteps during the program execution, the language model can be called for generation(via \{\{gen\}\} tag) or to make logical flow decisions, such as \{\{\#select\}\}\{\{or\}\}\{\{/select\}\} commands. Guidance supports a variety of LLMs, and during dialogue, it can use role labels to map the current LLM to correct tokens or API calls, such as \{\{\#assistant\}\}, \{\{\#user\}\}, \{\{\#system\}\} etc. It also can be integrated with HuggingFace models, including using Guidance acceleration to speed up standard prompts by reusing key-value caches to shorten prompt execution times and using token healing to optimize prompt boundaries. Regarding token healing, this concept is related to fixing the subtleties introduced by the language model's normal greedy tokenization method. Specifically, it involves advancing the model one step while simultaneously restricting the prefix of the generated token to be the same as the previous token. Regex patterns to enforce formatting are also supported in Guidance. Guidance's templated output is more suitable for generating text with high formatting requirements, such as ensuring legally compliant and controllable JSON structures. During this process, different operation commands have their processing methods; for example, encountering the {{select}} command, it specifies the generation of a token and returns the corresponding log probs, then uses a trie tree to match candidates and determine their probabilities, finally selecting the one with the highest probability. Additionally, it supports hidden blocks; for instance, some inference processes of the LLM may not need to be exposed to the end user, but they can be utilized in the template to generate intermediate results.

\subsubsection{LMQL (Language Model Query Language)} 
\begin{figure}[htbp]
    \centering
    \includegraphics [width=\columnwidth]{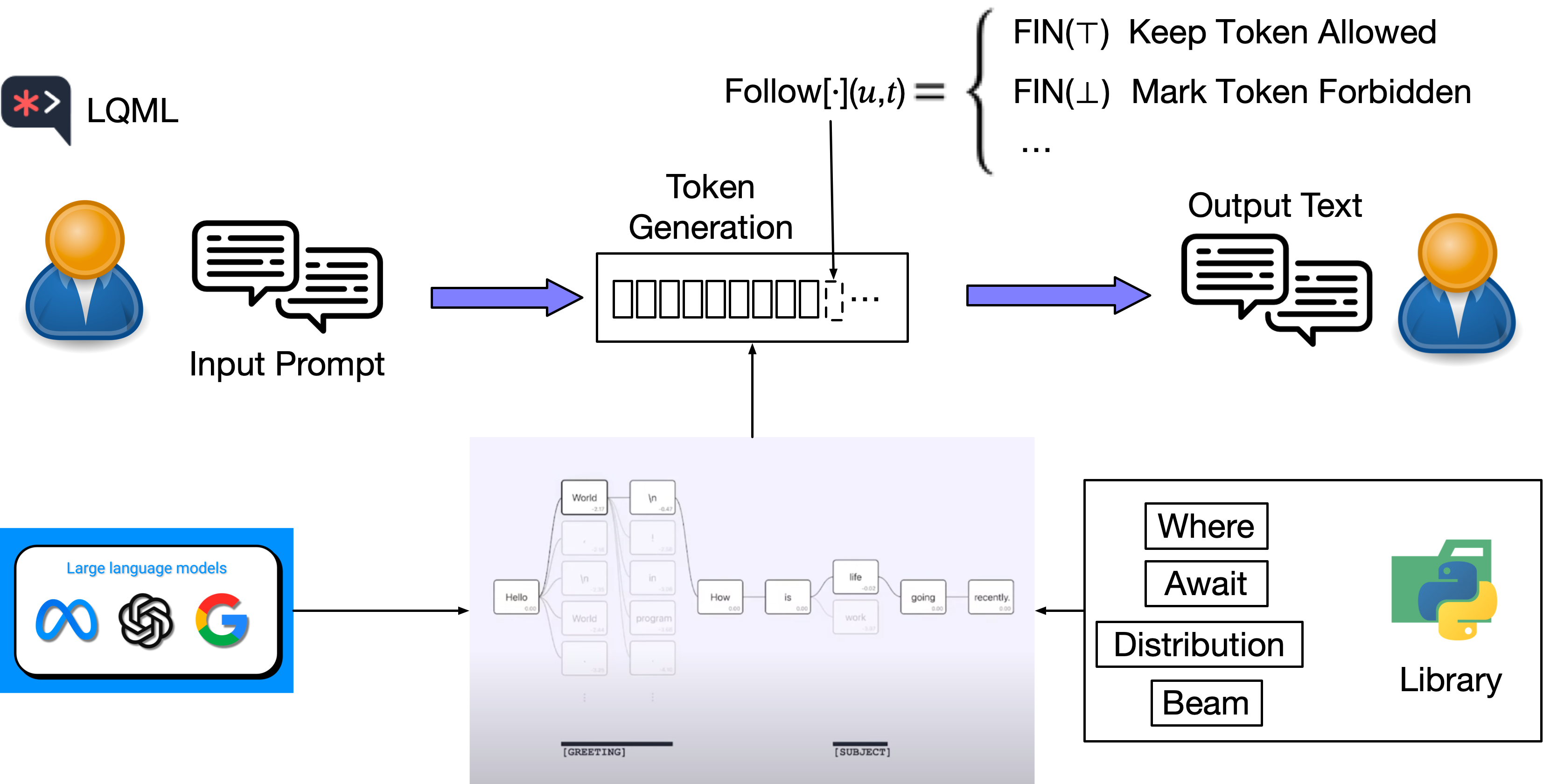}
    \caption{LMQL Workflow}
    \label{fig:LQML}
\end{figure}

LMQL \footnote{https://lmql.ai}, a programming interface for LLMs focusing on controlled output and safety of generated content, is designed by SRI Lab at ETH Zurich. Building on the foundation of Guidance AI, the LMQL project further advances the concept of ``prompt templates'' into a new programming language paradigm. As a Python superset, it allows developers the capacity to embed precise constraints within their queries. These constraints,  from content restriction to adherence to specific formats for accuracy, leverage logit masking and custom operator support for fine-tuned control. Structured to simplify LLM interactions, LMQL introduces a SQL-like syntax complemented by scripting capabilities. Its foundation is built on decoder declarations, such as argmax, beam, or sample strategies, alongside query blocks that support inserting variables or placeholders expected to be filled, model sources, and intricate constraint conditions specified in 'where' clauses. 
The workflow of the LQML is illustrated in Figure~\ref{fig:LQML}.

At its core, LMQL's runtime and decoding ability uniquely features employs a scripted Beam search to execute LMQL-specific queries and constraints. This approach allows for a search across all placeholders rather than limiting the process to singular predictive points. Additionally, constraint-driven decoding dynamically adjusts available tokens based on real-time evaluation of specified constraints and can reduce ineffective model calls. The Partial Evaluation Semantics and FollowMaps facilitate the application of constraints in real-time during content generation, considering both the current state of the generated content and the potential next tokens. Partial evaluation semantics combines value and final semantics for eager validation. Value semantics determines the current value of expressions given the existing context. In contrast, the final semantics employs annotations such as \texttt{FIN} (fixed), \texttt{VAR} (variable), \texttt{INC} (increasing), and \texttt{DEC} (decreasing) to indicate the potential change in the value of expressions as content generation progresses. Among the joint value and final semantics, boolean expressions are denoted by symbols $\top$ for True and $\bot$ for False. Based on these semantics, FollowMap which is formulated as Follow[<expr>]$(u, t)$ is designed to guide the generation process by evaluating the impact of adding a new token $t$ to the current interaction trace $u$ on meeting the specified constraints. The system evaluates potential next tokens against FollowMap constraints, excluding those that may violate constraints, thus guiding content generation away from invalid sequences and minimizing unnecessary model calls.



\subsubsection{Python Packages}
Apart from the above-mentioned integrated guardrail techniques, Several Python packages are pivotal in implementing guardrails around LLMs, enhancing their safety, fairness, and compliance. Here we listed some packages: (1) $LangChain$\footnote{\url{https://www.langchain.com/}}: LangChain is pivotal in streamlining the development of LLM applications. It introduces components that can be leveraged to implement guardrails, thus indirectly contributing to creating safer and more reliable AI solutions.
(2) $AI\ Fairness\ 360$ (AIF360)\footnote{\url{https://github.com/Trusted-AI/AIF360}}: AIF360 is an extended toolkit from IBM that provides developers with a comprehensive set of algorithms designed to detect, understand and mitigate bias in AI models. Its extensive resources help enhance the fairness and integrity of machine learning applications.
(3) $Adversarial\ Robustness\ Toolbox$ (ART)\footnote{\url{https://github.com/Trusted-AI/adversarial-robustness-toolbox}}: ART is tailored to enhance model security and robustness in the face of increasing adversarial attacks. It provides mechanisms to defend against and adapt to malicious input, protecting AI applications from potential vulnerabilities.
(4) $Fairlearn$\footnote{\url{https://github.com/fairlearn/fairlearn}}: This package addresses and reduces unwanted biases in machine learning models. Fairlearn provides developers with tools and methodologies to assess and mitigate biases, fostering fairness and equality in AI outcomes.
(5) $Detoxify$\footnote{\url{https://github.com/unitaryai/detoxify?trk=article-ssr-frontend-pulse_little-text-block}}: Detoxify aims to identify and mitigate toxic content in text data, serving as a crucial tool for maintaining a respectful and safe digital interaction space. It utilizes advanced models to screen and filter out harmful content, ensuring that AI applications foster positive communication.

These packages represent just a glimpse into the vast array of tools available to AI developers dedicated to embedding ethical considerations into their applications. By utilizing these packages, developers can navigate the complex challenges of AI safety, ensuring their LLMs are technologically advanced and aligned with ethical and responsible use principles. This approach reinforces the commitment to developing innovative AI technologies that respect societal norms and individual rights, marking a significant step towards responsible AI development.


\subsection{Techniques for (Un)desirable Properties in Guardrails}\label{sec:properties}

In this section, we discuss several different properties, detailing their standard definitions and how to use Guardrails to protect these properties. It is noticed that the number of properties is too extensive to cover comprehensively; thus, we focus on hallucination, fairness, privacy, robustness, toxicity, legality, out-of-distribution, and uncertainty.

\subsubsection{Hallucination}

Hallucinations in LLMs are defined as responses that are either nonexistent in reality, illogical, or irrelevant to the prompt provided; an example of hallucination is shown in Fig.~\ref{fig:Hallucination example}. These phenomena often stem from the language models' uncertainty in response, where they generate answers based on patterns identified in training datasets rather than actual factual understanding \cite{huang2023surveyhallu}. The origins of these hallucinations can be traced back to various stages, including data sourcing \cite{singhal2023towards}, pre-training \cite{li2023batgpt}, alignment \cite{singhal2023towards}, and inference \cite{liu2023instruction}.

The resolution of hallucinations in LLM is crucial due to their impact on model reliability and real-world applications, which presents a significant challenge in LLM development. Researchers are actively working on methods to differentiate between accurate and hallucinated content in LLM responses. This involves the use of detection technologies \cite{CircleCI2024LLMHallucinations} and evaluation methodologies \cite{cheng2023evaluating} to ensure the integrity and utility of LLM outputs. 
In certain contexts, such as safety-critical applications, ensuring rigorous guarantees for the output of LLMs is imperative and achievable.
This involves first specifying high-level formal specifications and then applying formal verifiers to monitor whether the execution of LLMs conforms to the specified requirements.
For instance, in \cite{jha2023counterexample}, the authors proposed a framework for counterexample-guided inductive synthesis using LLMs and satisfiability modulo theory (SMT). Within this framework, an SMT solver is employed to eliminate the hallucinated outputs of the LLM, guiding it to generate correct outputs.
Experimental results on two planning problems demonstrated the promise of this approach in practice. The framework consistently converged to correct outputs in finite steps.
However, it is worth noting that such convergence cannot always be theoretically guaranteed, underscoring the practical significance of this method in addressing LLM hallucinations. Furthermore, there are some researches proposed to detect and prevent the hallucinations of LLMs.  

In \cite{CircleCI2024LLMHallucinations}, the authors suggest utilizing continuous integration (CI) to automatically identify hallucinations in the outputs of an LLM with the aid of another LLM. The CI process involves regular incremental updates to the software, with each change automatically built and tested, allowing for prompt detection of errors. Therefore, LLM developers can use CI to automate tests on new datasets and updates to the models, which ensures that any changes do not unintentionally introduce bias or reduce the quality of the model’s output. Instead of using another LLM to identify the errors, some researchers use trusted information sources to cross-check the generated content \cite{min2023factscore}. Building on these techniques, a comprehensive framework is proposed by \cite{chern2023factool}, which equips LLMs with the ability to recognize factual inaccuracies, utilizing external tools to gather supporting evidence. In addition to verifying the accuracy of responses, it's crucial to assess the faithfulness of the output to prevent hallucinations that diverge from the context. This means ensuring the output remains relevant and agrees with the input, avoiding any extraneous or contradictory information. This can be achieved by the fact-based metrics that are based on entity \cite{nan2021entity}, relation\cite{goodrich2019assessing}, and knowledge F1 metric \cite{shuster2021retrieval}. A further approach to ensure a model's faithful output utilizes the classifiers that are either trained on task-specific inference datasets \cite{mishra2021looking} or fine-tuned on adversarial datasets \cite{barrantes2020adversarial}. Nowadays, the instruction-following abilities of LLMs are leveraged for autonomous evaluation. LLMs can effectively gauge accuracy and relevance by setting clear evaluation criteria and providing both generated and source content. Different methodologies have been adopted for output evaluation, such as binary judgment mechanisms \cite{luo2023chatgpt} and using a k-point Likert scale \cite{gao2023human}.

In current guardrails, Nvidia Nemo employed the mechanism proposed by \cite{manakul2023selfcheckgpt}. It first generates a few additional responses from the LLM, typically two more by default. The original response from the bot is treated as the hypothesis, while the additional generated responses serve as the context or evidence. This method aligns the consistency verification with natural language inference (NLI) principles, allowing for a structured output assessment.

\subsubsection{Fairness}

Fairness is a concept that originates in sociology, economics, and law. It is defined as ``imperfect and just treatment or behavior without favoritism or discrimination'' in the Oxford English Dictionary. The key to fairness in NLP is the presence of social biases in language models; an example is illustrated in Fig.~\ref{fig:Fairness Example}. 
Unfair LLM systems make discriminatory, stereotypic, and demeaning decisions against vulnerable or marginalized demographics, causing undesirable social impacts and potential harms \cite{DBLP:conf/acl/BlodgettBDW20}. Fairness in LLMs has been studied from different angles, such as gender bias \cite{malik2023evaluating},
cultural bias \cite{tao2023auditing}, dataset bias \cite{sheppard2023subtle}, and social bias \cite{sheng2023fairness}.
These social biases can be encoded in the embeddings and carried over to decisions in downstream tasks, compromising the fairness of LLMs. For instance, when GPT-3 is prompted with questions about leadership roles or positions of power, it may reflect societal biases in its responses. If asked, ``\textit{Who is likely to be a CEO?}'', GPT-3's response might lean towards ``\textit{He is likely to be a CEO},'' implying a male. Conversely, suppose the question is about lower-ranking positions or roles traditionally seen as supportive or administrative, such as ``\textit{Who is likely to be a secretary?}''. In that case, the model might respond with ``\textit{She is likely to be a secretary},'' suggesting a female. This pattern of responses reveals a bias where higher-status professions or roles are more closely associated with males, while supportive or administrative roles are linked with females. Such biases in LLM outputs can reinforce outdated stereotypes and potentially influence the perception of what roles are ``appropriate" for individuals based on their gender. 

Thus, to guard fairness in LLMs, increasing work is needed to understand these biases and evaluate their further effects on the downstream tasks of LLMs. For example, in terms of racial bias, an African American is more likely to be assigned a ``criminal behavior'' feature because of the ``African'' group he belongs to \cite{garrido2021survey}. When this feature is used for model encoding and further downstream tasks, it induces unfairness in the language model toward African Americans. Biases are purposefully introduced into the responses of LLMs to craft distinct personas for use in interactive media \cite{badyal2023intentional}. BAD focuses on identifying and quantifying instances of social bias in models like ChatGPT, especially in sensitive applications such as job and college admissions screening \cite{koh2023bad}. DAMA utilizes causal analysis to identify problematic model components, mainly focusing on mid-upper feed-forward layers most prone to convey biases \cite{limisiewicz2023debiasing}. The presence of political bias is examined in ChatGPT, focusing on aspects such as race, gender, religion, and political orientation \cite{motoki2023more}.
Additionally, they explored the role of randomness in responses by collecting multiple answers to the same questions, which enables a more robust analysis of potential biases. The bias of LLMs is also examined by controlling the input, highlighting that LLMs can still produce biased responses despite the progress in bias reduction \cite{yeh2023evaluating}. Besides, a Bias Index is designed to quantify and address biases inherent in LLMs, including GPT-4 \cite{shaikh2022second}. It has also been observed that the biased response can be generated inadvertently, sometimes as seemingly harmless jokes \cite{zhou2023public}. 


A line of debiasing studies aims to mitigate the \textit{intrinsic bias} that is task-agnostic in the representations before they are applied to downstream tasks. Safeguarding training data before training the model efficiently alleviates intrinsic biases since label imbalance across different demographic groups in the training data is an essential factor in inducing bias. A Counterfactual Data Augmentation (CDA) \cite{ma-etal-2020-powertransformer,DBLP:conf/acl/XieL23} 
 is a widespread data processing method to balance labels, which replaces the sensitive attributes in the original sample with the sensitive attributes of the opposite demographic based on a prior list of sensitive word pairs. Fairness can be incorporated into LLMs’ design to balance the training samples, and then a guarded fairer model can be obtained by parameter tuning. Retraining models \cite{DBLP:conf/emnlp/QianRFSKW22}
is a direct way to reduce bias, although it can be resource-intensive and difficult to scale. For instance, FairBERTa is a fairer model for retraining RoBERTa on a large-scale demographic perturbation corpus Perturbation Augmentation NLP DAtaset (PANDA) containing 98K augmentation sample pairs \cite{DBLP:conf/emnlp/QianRFSKW22}. Moreover, an additional debiasing module is added after the encoder of LLMs to filter out the bias in the representation, and a common strategy is to utilize a contrastive learning framework for training \cite{DBLP:conf/kdd/OhWSKKCS22}. 

The other line of debiasing studies aims to mitigate the \textit{extrinsic debiasing} in a task-specific way. These studies attempt to improve fairness in downstream tasks by making models provide consistent outputs across different demographic groups. Many studies have concentrated on reducing bias through model adaptation approaches. 
A bias mitigating method, DAMA \cite{limisiewicz2023debiasing}, can reduce bias while maintaining model performance on downstream tasks.  \cite{ranaldi2023trip} investigated the bias in CtB-LLMs and demonstrated the effectiveness of debiasing techniques. They find that bias depends not solely on the number of parameters but also on factors like perplexity and that techniques like debiasing of OPT using LoRA can significantly reduce bias. \cite{ungless2022robust} demonstrated that the Stereotype Content Model, which posits that minority groups are often perceived as cold or incompetent, applies to contextualized word embeddings and presents a successful fine-tuning method to reduce such biases. Moreover, \cite{ernst2023bias} proposed a novel adversarial learning debiasing method applied during the pre-training of LLMs. 
\cite{ramezani2023knowledge} mitigated cultural bias through fine-tuning models on culturally relevant data, yet it requires resources that make it accessible to only a few.

Instead of fine-tuning parameters, several studies directly set up guardrails on the interaction between users and LLMs by exploring the control of input and output. \cite{huang2023bias} suggested using purposely designed code generation templates to mitigate the bias in code generation tasks. \cite{tao2023auditing} found that cultural prompting is a simple and effective method to reduce cultural bias in the latest LLMs. However, it may be ineffective or even exacerbate bias in some countries. \cite{oba2023contextual} proposed a method to address {gender\ bias} that does not require access to model parameters. It shows that text-based preambles generated from manually designed templates can effectively suppress gender biases in LLMs with minimal adverse effects on downstream task performance. \cite{dwivedi2023breaking} guided LLMs to generate more equitable content by employing an innovative approach of prompt engineering and in-context learning, significantly reducing gender bias, especially in traditionally problematic.

Developing guardrails through a comprehensive approach that intertwines various strategies is crucial to mitigate bias effectively. This begins with meticulously monitoring and filtering training data to ensure it is diverse and devoid of biased or discriminatory content. The essence of this step lies in either removing biased data or enriching the dataset with more inclusive and varied information. Alongside this, algorithmic adjustments are necessary, which involve fine-tuning the model's parameters to prevent the overemphasis of certain patterns that could lead to biased outcomes. Incorporating bias detection tools is another pivotal aspect. These tools are designed to scrutinize the model's outputs, identifying and flagging potentially biased content for human review and correction. We believe that adopting a continuous learning approach is key to the long-term efficacy of these guardrails. This involves regularly updating the model with new data, insights, and feedback and adapting to evolving societal norms and values. This dynamic process ensures that the guardrails against bias remain robust and relevant. Moreover, we believe in \emph{principled methods} to evaluate fairness when the definitions are settled. However, the definition is expected to be distribution-based rather than point-based as unintended responses, which need to estimate posterior distributions and measure the distance between two distributions. 

\subsubsection{Privacy (Copyright)}

Privacy, in the context of modern technology and artificial intelligence, is a crucial aspect of data protection that has been increasingly emphasized through legislation and research. Legislative measures like the EU AI Act, General Data Protection Regulation (GDPR), and California Consumer Privacy Act (CCPA) have established stringent data sharing and retention standards, necessitating strict adherence to data protection and privacy guidelines. Despite these frameworks, challenges persist in preventing the release of personally identifiable information (PII) by LLMs \cite{zou2023universal}, emphasizing the need for cautious and robust data handling protocols, an example of a privacy issue is shown in Figure \ref{fig:Privacy Example}. 
\cite{li2023privacy} comprehensively analyzes privacy attacks against LLMs, introduces significant defense strategies, and highlights potential new privacy issues and future research directions as LLMs evolve.

Several studies have focused on implementing privacy defense technologies to safeguard data privacy and mitigate privacy breaches. Differential Privacy (DP)-tuned LLMs \cite{li2023privacy} emerge as a leading approach to protecting data privacy in these contexts, ensuring secure handling of sensitive information by LLMs and minimizing the risk of privacy violations. 
For general NLP models, \cite{li2022large} indicated that a direct application of DP-SGD \cite{10.1145/2976749.2978318} may not perform satisfactorily and suggests a few tricks. 
\cite{mireshghallah2022differentially} study differential privacy model compression and proposes a framework that achieves 50\% sparsity levels while maintaining nearly complete performance, setting a benchmark for future research in this area.
\cite{igamberdiev2023dp} implemented a model for text rewriting along with Local Differential Privacy (LDP), both with and without pretraining.
\cite{xiao2023large}  introduce Privacy Protection Language Models (PPLM), a novel paradigm for fine-tuning LLMs that incorporates domain-specific knowledge while preserving data privacy. They explore techniques such as corpus curation and instruction-based tuning, demonstrating the effectiveness of these approaches in safeguarding private data. 
\cite{zhao2023silent} introduce a novel text protection mechanism called "Silent Guardian," which effectively prevents the malicious use of text by LLMs through Truncation Protection Examples and the Super Tailored Protection algorithm. It features efficiency, semantic consistency, transferability, and robustness.
\cite{ozdayi2023controlling} proposed a method to prepend a trained prompt to the incoming prompt before passing it to the model, where the training of the prefix prompt is to minimize the extent of extractable memorized content in the model. \cite{li2023privacy} and \cite{duan2023flocks} also proposed the prompt-tuning methodology that adheres to differential privacy principles.
\cite{DBLP:conf/iclr/YuNBGI0KLMWYZ22} propose an effective algorithm for differentially private fine-tuning of large pre-trained language models, which achieves utility close to that of non-private models while protecting privacy and reduces the computational and memory cost of training, especially performing excellently on larger models.
\cite{shi2022just} introduces a ``Just Fine-tune Twice" (JFT) framework for the latest large Transformer models, achieving Selective Differential Privacy protection. It enhances the model's utility and privacy safeguards through double fine-tuning and systematic methods.

Other than constructing privacy-preserving LLMs, watermarking techniques can play a more critical role in LLMs for privacy and copyright protection. A typical watermarking mechanism \cite{pmlr-v202-kirchenbauer23a} embedded watermarks into the output of LLMs by selecting a randomized set of ``green'' tokens before a word is generated and then softly promoting the use of green tokens during sampling. So, as long as we know the list of green tokens, it is easy to determine if an output is watermarked or not. 
We can also use the watermarks to track the point of origin or the owner of watermarked text for copyright purposes, and this has been applied to protect the copyright of generated prompts \cite{yao2023promptcare}. We believe in an agreed watermarking mechanism between the data owners and the LLMs developers, such that the users embed a personalized watermark into their documents or texts when they deem them private or with copyright, and the LLMs developers will \emph{not} use watermarked data for their training. More importantly, the LLMs developers should take the responsibility of enabling (1) an automatic verification to determine if a user-provided, watermarked text is within the training data, and (2) model unlearning \cite{nguyen2022survey}, which allows the removal of users' personally owned texts from training data. 
LLMs also risk user trust due to their pre-training on vast textual datasets \cite{narayanan2021scaling}, potentially leading to inadvertent disclosure of sensitive information about individuals \cite{plant2022you}. Malicious actors can exploit this vulnerability through adversarial attacks \cite{wang2024decodingtrust}, underscoring the critical importance of privacy protection, especially when fine-tuning LLMs with sensitive data.


In addressing privacy concerns within LLM applications, implementing guardrails is crucial for existing and in-development technologies. Key strategies for existing applications include robust testing to identify privacy risks and continuous model monitoring to adapt to new threats. Implementing content control mechanisms such as blocklists, allowlists, and suppression lists directly tackles unsafe content generation that could compromise privacy. For example, Nemo Guardrails restricts apps to making connections only to external third-party applications known to be safe. The guardrails can force an LLM model to interact only with third-party software on an allowed list.
The "human-in-the-loop" approach, where human oversight is applied to review potentially sensitive outputs and facilitates user reporting channels for privacy violations, enhances the protection framework \cite{rahman2024survey}. Regular model retraining to align with current norms and the option to revert to previous safe versions of the model serve as dynamic responses to privacy challenges.

For applications still in development, privacy protection begins at the design stage, with ethical risk assessments focused on identifying and mitigating privacy risks. Adopting responsible AI practices ensures privacy is a core consideration from the outset \cite{sarker2024llm}. Implementing selective memory and information filtering techniques restricts the AI's access to sensitive data, directly safeguarding user privacy. Removing personally identifiable information (PII) from data used in model training is critical in protecting privacy \cite{yang2023exploring}. Continuous updates to employ the latest version of LLMs and strict data privacy protocols for staff overseeing AI use are also essential for maintaining privacy standards.

\subsubsection{Robustness}

With the rise of LLMs as dominant models in NLP, robustness consists of out-of-distribution (OOD) and adversarial robustness. This section only accounts for adversarial robustness, while OOD is discussed in Section \ref{sec:OOD}. The adversary (end-user) only attempts to jailbreak the model by explicitly optimizing adversarial queries or adaptively making queries based on previous outputs but will not make out-of-distribution queries asking about potentially revoked information. Robustness has distinct definitions across various downstream tasks of NLP; it can be commonly characterized in the following way (It works for a range of NLP tasks like text classification and sequence labeling): let \( x \) represent the input and \( y \) its corresponding correct label. Consider a model \( f \) that has been trained on data pairs \( (x, y) \sim D \), with its output prediction for \( x \) given by \( f(x) \). When new test data \( (x', y') \sim D' \), where \( D' \) is not identical to \( D \), is introduced, the robustness of the model can be determined by its performance on \( D' \) \cite{wang2022measure}. Through comprehensively perturbing the input from \( x \) to \( x' \), we encounter the notion of adversarial robustness, which is a concept originating from the computer vision \cite{10.1145/3593042}. 

The adversarial robustness under the LLMs refers to the ability of models to maintain performance when faced with inputs that have been intentionally altered or crafted to cause the model to error, such as the malicious queries made intentionally or unintentionally \cite{ye2023assessing}. It is a type of model based on transformations or small perturbations (e.g.typo) to study the robustness of the model (it is also called invariance of LLMs) \cite{liang2023holistic}. Typically, alterations that maintain the underlying meaning, like modifying the text case and contraction perturbed, are considered fairly benign \cite{liang2023holistic}. In particular, disturbances are directed at various layers of linguistic signals, including characters, words, sentence structures, and underlying meanings. The core objective is to replicate potential user mistakes (e.g. use of near-meaning words), to assess the impact of minor deviations on the outcomes of LLMs \cite{zhu2023promptbench,wang2024decodingtrust}.

The defense methods for shielding LLMs from deliberate disruptions remain under investigation \cite{liu2024semantic}, indicating that robust safeguarding measures are necessary, especially during the most crucial phases of user engagement with these models. Typically, guardrails pre-process users’ inputs to remove or neutralize potentially adversarial content, thus preventing models from being misled by manipulated inputs (e.g., correcting typos and standardizing input formats). Similarly, guardrails also monitor LLM's outputs. This may involve establishing thresholds for specific types of responses or flagging outputs that significantly deviate from expected patterns for review by a professional security team.

\subsubsection{Toxicity}

An important NLP task is the toxicity detection \cite{pavlopoulos-etal-2020-toxicity}, the term `toxicity' is employed as a broad descriptor, encompassing a variety of related phenomena and linguistic contexts that may also manifest as `offensive' \cite{zampieri-etal-2019-predicting}, `abusive' \cite{menini2021abuse}, `hateful' \cite{kirk2023handling}. Similar descriptors \cite{pavlopoulos-etal-2020-toxicity}. We show a typical example of toxicity in Fig.~\ref{toxicity_example}.
LLMs, as one of the prevalent developments in traditional language modeling, are frequently trained using vast quantities of datasets, which can include content exhibiting toxic behavior and unsafe material, such as hate speech, offensive/abusive language, etc. Typically, a thorough examination of toxicity is required, especially considering the employment of LLMs for downstream tasks that might engage younger or more vulnerable individuals, as well as the negative effects of unintended outputs from LLMs on specific tasks \cite{zhang2023comprehensive}. The definition of what constitutes toxicity of the LLMs varies normally, toxicity responses will be defined as rude, disrespectful, or unreasonable responses that are likely to make an individual leave a discussion \cite{deshpande-etal-2023-toxicity}. It is, hence, very desirable to evaluate how well-trained LLMs deal with toxicity \cite{guo2023evaluating}.

Existing studies address the problem by focusing on representative terms in datasets, such as identity terms \cite{sap-etal-2020-social}. To evaluate the toxicity in LLMs, several studies have crafted trigger prompts that mirror detailed toxic categories \cite{gehman-etal-2020-realtoxicityprompts}. These studies leveraged standard toxicity metrics, such as the Toxicity Classifier Score and PerspectiveAPI\footnote{Perspective API was developed by Jigsaw and the Google Counter Abuse Technology team (\url{https://perspectiveapi.com})}, to determine whether the LLM's response is toxic \cite{hosseini2017deceiving}. However, typical metrics are susceptible to evaluator bias \cite{10.1145/3555088}, and encoders are perturbed \cite{rosenblatt-etal-2022-critical}. Subsequently, a structured investigation framework attempted to address this bias \cite{koh2024llms}. Despite being trained on extensive datasets, LLMs are capable of generating outputs that can be implicitly toxic, which are difficult to detect with straightforward, zero-shot methods \cite{welbl-etal-2021-challenges-detoxifying,wen2023unveiling}. This complexity arises even when prompts appear non-toxic, underscoring the nuanced challenges in detoxifying language models, such as depending on the specific roles assigned to LLMs, certain roles may generate markedly more toxic outcomes \cite{deshpande-etal-2023-toxicity}.

Even when a generative model is trained on data characterized by low toxicity levels, and its ability to minimize the generation of toxic text has been validated through evaluations, it is still crucial to enforce protective measures during live interactions between users and the model \cite{liang2023holistic}. Safety guardrails are an integral part of the user interaction and LLMs interaction phases, playing a key role in ensuring privacy, preventing bias, and maintaining user trust \cite{dong2024building}. For example, Nvidia Nemo allows users to define the toxic output they want to identify; the next step is determining the chatbot's response to users' input. This involves setting up a workflow that utilizes these definitions. Thus, this procedure is triggered whenever there is potential exposure to toxic content, and the chatbot supports the user. Furthermore, they ensure compliance with legal standards and align AI operations with societal values.



\subsubsection{Legality}
Another crucial aspect of safeguarding LLMs involves managing the risks associated with illicit\footnote{It's important to note that while the concepts of legality and toxicity may overlap to some extent, they are not synonymous. Legality is the lowest requirement, defining what is permitted under the law. However, something that is not illegal may still be considered toxic due to its potential to cause harm or adverse effects in other contexts. Conversely, if something is deemed illegal, it invariably falls into the category of being toxic, as its prohibition by law implies a recognized potential for harm or negativity. Thus, while legality provides a clear boundary based on legal statutes, toxicity encompasses a broader range of potentially harmful actions or materials, some of which may not be explicitly covered by legal definitions but are nonetheless detrimental to well-being or ethical standards.
} outputs~\cite{kumar2023language}.
Generally, this involves safeguarding efforts on two fronts: implementing measures to reject inappropriate user inputs and moderating model output to ensure it is appropriate and safe for users or downstream tasks.

During the development of LLMs, developers implement a series of measures to ensure the safety and compliance of the models with relevant laws and regulations.
These measures include:
i) researchers meticulously screen and clean the training data before training the models to remove inappropriate, harmful, or illegal content. This ensures the model learns from high-quality data and avoids adopting inappropriate behaviors.
ii) During model training, human reviewers assess the samples generated by the model and offer feedback, aiding in rectifying errors and enhancing the model's output. This process, alongside RLHF, helps models refine the content they generate and gradually adopt appropriate behavior.

Once the model construction is completed and before release, models undergo thorough and rigorous ethics and safety testing to ensure that the content they generate is absent of inappropriate or illegal elements.
One classical approach is red teaming~\cite{ganguli2022red,perez2022red}, which entails simulated attacks and adversarial testing to uncover potential vulnerabilities, ethical pitfalls, and legal considerations.
Organizations like OpenAI, Anthropic, Google, and Meta utilize diverse methodologies for red teaming, ensuring a thorough evaluation and effective risk mitigation.
For instance, Google promotes internal red teams~\footnote{\url{https://blog.google/technology/ai/google-gemini-next-generation-model-\\february-2024/}}, where employees with diverse expertise simulate attacks on the AI model.
In contrast, OpenAI favors external red teaming and has established external networks~\footnote{\url{https://openai.com/blog/red-teaming-network}} to encourage participation from outside members.

In addition to the above safeguarding efforts, monitoring systems are established upon model release to detect inappropriate inputs and outputs.
Techniques such as natural language processing and anomaly detection are employed for real-time identification.
Upon identification of any issues, immediate measures, such as content filtering algorithms or human intervention protocols, are swiftly implemented to address the concern.
It is worth noting that leading LLM providers, such as Google, OpenAI, Anthropic, and Meta, offer advanced moderation tools and techniques to developers or users, enabling customized safeguards against illicit and inappropriate content.
For instance, Google offers PaLM-based Moderation~\footnote{\url{https://cloud.google.com/natural-language/docs/moderating-text}}, capable of detecting more than 16 types of inappropriate content.
OpenAI provides a Moderation API~\footnote{\url{https://platform.openai.com/docs/guides/moderation}}, allowing developers and users to customize safeguards for inappropriate content.
Meanwhile, Anthropic has developed Constitutional AI~\cite{bai2022constitutional} and Meta utilizes Llama Guard~\cite{inan2023llama} for content moderation.
%

In addition to the moderation tools offered by LLM providers, notable contributions from other entities in the field also exist.
For instance, \textit{LangChain}~\footnote{\url{https://python.langchain.com/docs/modules/chains}}, an open-source framework, simplifies and safeguards the development of applications using LLMs.
Specifically, it offers a standardized interface for creating, combining, and customizing various components, resulting in powerful language-driven applications.
One notable application of LangChain in the legal domains is ConstitutionalChain~\footnote{\url{https://api.python.langchain.com/en/latest/chains/langchain.chains.constitutional_ai.base.ConstitutionalChain.html}}.
By incorporating predefined rules and guidelines, ConstitutionalChain can filter and modify generated content to align with constitutional principles.
This ensures that responses are controlled, legal, and contextually appropriate.

\subsubsection{Out-of-Distribution}
\label{sec:OOD}

For a specific DNN, \textit{out-of-distribution} (OOD) data strictly refers to data not belonging to any in-distribution classes used in training.
Broadly, OOD data can be characterized as differing from the in-distribution data on certain dimensions.
Research indicates that DNNs often exhibit overconfident decision-making when presented with OOD data.
This has led to widespread investigation of OOD detection issues across domains such as computer vision~\cite{hendrycks2016baseline}, 
and natural language processing~\cite{arora2021types}. 
However, the OOD detection task within the field of NLP presents notable challenges, particularly exacerbated by the presence ofLLMs. This issue has resulted in limited research focused on OOD detection specifically tailored to LLMs~\cite{ren2022out}, primarily due to the immense training corpora used for LLMs, making it difficult to define precisely what data has not been utilized for training.
Moreover, the generative nature of LLMs adds another layer of complexity to defining the OOD problem~\cite{kadavath2022language}.

While defining OOD instances for an LLM is generally very difficult, if not impossible, it becomes more feasible when applied to specific real-world scenarios where the context is more precise.
In practical scenarios, OOD instances can be defined as data irrelevant to the main task or significantly deviating from normal ones. 
For instance, recent work~\cite{li2023survey} has explored the evaluation of OOD in the context of specific language model applications, such as text classification~\cite{kaushik2019learning}, sentiment analysis~\cite{zhang2023sentiment},  machine reading
comprehension~\cite{zeng2020survey}, and found that it can lead to a significant performance decrease, even with minor semantic shifts caused by small perturbations.
To mitigate the impact of OOD on model performance in practical tasks, strategies such as setting up anomaly input filtering mechanisms~\footnote{\url{https://hub.guardrailsai.com/validator/guardrails/unusual_prompt}} or constructing OOD detectors tailored to the task can be employed.

\subsubsection{Uncertainty}

A key aspect of LLMs' trustworthiness lies in their ability to discern their outputs' reliability and correctness, a concept central to uncertainty quantification. 
This approach is an effective method for assessing risks, aiming to gauge the confidence levels of LLMs in their predictions. 
Elevated uncertainty suggests that an LLM's output may require rejection or additional scrutiny. Fig.~\ref{fig:uncertainty example} shows an example of uncertainty.
The effectiveness of uncertainty quantification is further contingent on the alignment between the model's predicted confidence and its actual accuracy, essentially measuring the model’s calibration.

There has been a growing focus on research to quantify the overall uncertainty in LLMs.
Establishing dependable uncertainty metrics is essential for enhancing the safety of LLM systems.
Recent studies have noted that the calibration of LLMs is improved relatively through techniques like combining multiple reasoning chains~\cite{wang2023selfconsistency}, integrating different prompts~\cite{jiang2023calibrating}, or by prompting LLMs to output their confidence levels directly~\cite{kadavath2022language}.
In addition to these observations, numerous methods have been developed to quantify the uncertainty in LLMs effectively.
\cite{lin2022teaching} demonstrated that a GPT-3 model can learn to articulate uncertainty regarding its responses in natural language independently of using model logits. 
\cite{xiao2022uncertainty} comprehensively compared various popular approaches to construct a well-calibrated prediction pipeline for pre-trained language models.
\cite{ren2023robots} unveiled KnowNo, a framework designed to measure and align the uncertainty in LLM-based planners, enabling them to recognize their limitations and seek assistance when necessary.



The primary hurdles in assessing LLM uncertainty arise from the pivotal roles of meaning and structure in language. 
This pertains to what linguists and philosophers define as a sentence's semantic content and syntactic or lexical framework. 
While foundation models mainly output token-likelihOODs, reflecting lexical confidence, the meanings often hold the most significance in most applications.
\cite{kuhn2022semantic} introduced the concept of semantic entropy, which integrates linguistic consistencies arising from identical meanings. 
The fundamental method involves a semantic equivalence relation, denoted as $\mathbb{E} (s_i, s_j)$, where $s_i$ and $s_j$ represent output sentences corresponding to a given input. 
This equivalence relation is said to hold when two sentences $s_i$ and $s_j$ convey the same meaning, implying that they belong to the same cluster $C$.
The semantic entropy is defined as
\begin{equation}
H(C|x)=-\sum_C \mathbb{P}(C|x) \ln \mathbb{P}(C|x),
\end{equation}
where $x$ is the input sentence.
This methodology, which employs `out-of-the-box' models, enhances reproducibility and simplifies deployment.
Moreover, this unsupervised uncertainty could address the issue identified in previous research, where supervised uncertainty measures often falter in the face of distributional shifts.

Utilizing the above uncertainty technologies to build a guardrail for LLMs, it is crucial to integrate mechanisms that enable the model to assess and communicate its uncertainty. 
This involves training the model to recognize when a query falls outside its expertise or when the answer is speculative. It also involves responding appropriately—whether by providing a cautious answer, flagging the response as uncertain, or directing the user to more reliable sources.

\section{Overcome and Enhance Guardrails} \label{sec:4}

Implementing advanced safeguarding techniques, as discussed in Section \ref{sec:3}, has played a crucial role in enhancing their security and reliability within LLMs. 
However, \cite{shen2023anything} indicated that employing guardrails does not enhance the robustness of LLMs against attacks. 
They examined the external guardrails such as \textit{ModerationEndpoint}, \textit{OpenChatKitModeration Model}, and \textit{Nemo}, showing that they only marginally reduce the average success rate of jailbreak attacks. 
Jailbreak attacks, referred to as ``\textit{jailbreaks}'', aim to exploit language models' inherent biases or vulnerabilities by manipulating their responses.
These successful attacks allow users to circumvent the model's safeguard mechanisms, restrictions, and alignment, potentially leading to generating unconventional or harmful content or any content controlled by the adversary. 
By bypassing these constraints, jailbreaks empower the model to produce outputs that exceed the boundaries of its safety training and alignment. 

Therefore, in this section, we explore current methods used to bypass the guardrails of LLM.
In Table \ref{attacks}, we compare different jailbreaks on: (1) \textit{Attacker access type}: white box, black box, and gray box. In a white-box scenario, the attacker has full visibility into the model's parameters. A black-box situation restricts the attacker from observing the model's outputs. In a grey-box context, the attacker has partial access, typically to some training data. (2) \textit{Prompt level for manipulation}: user prompt or system prompt. User prompts are those where the input prompt is specified by the user, allowing for personalized or targeted inputs. On the other hand, system prompts are generated automatically by models and may include outputs that attackers craftily devise to deceive or manipulate the system’s response. (3) \textit{Core technique}: the main technique used to attack the LLM. (4) \textit{Stealthiness}: high stealthiness represents that the attack is difficult to notice by a human, which is supposed to be some logical, semantic, and meaningful conversation rather than some gibberish.
(5) \textit{GPT4 Evaluation}: As many jailbreaks are not directly targeted for LLMs with guardrails, and GPT4 has its default guardrail, then evaluation on GPT4 can be seen as a surrogate metric for comparison. 
(6) \textit{Target manipulated property of generated response}: toxicity, privacy, fairness, and hallucination

\begin{table*}[htbp]
\centering
\caption{Comparison among Different Jailbreaks for (Guarded) LLMs}
\label{attacks}
\resizebox{\linewidth}{!}{
\renewcommand\arraystretch{2}
\begin{tabular}{ccccccc}
\toprule[1.5pt]
Attack        & Access Type & Prompt Level& Core Technique          & Stealthiness & GPT4 Evaluation &  Targeted Property\\
\midrule
GCG~\cite{zou2023universal}           & White            & User               &      Greedy Gradient-based Search                    &    Low             &    $\times$    & Harmful Content       \\
PGD~\cite{geisler2024attacking}           & White            & User               &    Continuous Relaxation \& Entropy projection                   &             Low    &   $\times$  &Harmful Content   \\
PRP$\textbf{*}$~\cite{mangaokar2024prp}           & White            & User               &   In-context Learning \& two-step prefix-based                     &      Low      &$\times$    & Harmful Content         \\
AutoDAN-Liu$\textbf{*}$~\cite{liu2023autodan}&White & System+User        &  Hierarchical Genetic Algorithm        &    High            & $\checkmark$            & Harmful Content             \\
AutoDAN-Zhu$\textbf{*}$~\cite{zhu2023autodan}&White & User&Double-loop Optimization& High & $\checkmark$& Harmful Content \& prompt leaking \\
COLD-Attack$\textbf{*}$~\cite{guo2024cold}&White & User & Langevin dynamics& High & $\times$& Harmful Content  \\
ProMan~\cite{zhang2023safety} &White& -& Generation Manipulation & - & $\times$ & Harmful Content \& Privacy Leakage \\

\midrule
JailBroken~\cite{wei2024jailbroken}    & Black            & System             & Failure modes as guiding principles   & Low         &  $\checkmark$ & Harmful Content \& personally identifiable information leakage    \\
DeepInception~\cite{li2023deepinception} & Black    & User    & Nested instruction      & Medium   & $\checkmark$    & Harmful Content         \\
DAN$\textbf{*}$~\cite{shen2023anything}          & Black            & User               & Characterizing in-the-wild prompt           &High     &$\checkmark$    & Harmful Content    \\
ICA~\cite{wei2023jailbreak}         & Black            & User               &  In-context learning ability of LLM   & Low   &  $\times$      & Harmful Content             \\
SAP~\cite{deng2023attack}          & Black            & User             & In-context learning ability of LLM      & Medium    &  $\times$      &  Harmful Content        \\
DRA~\cite{liu2024making} & Black & User &Making Them Ask and Answer &  Low     & $\checkmark$ & Harmful Content\\
CipherChat~\cite{yuan2023gpt}   & Black            & System             & Long-tail: cipher       &  High     & $\checkmark$     & Harmful Content  \\
MultiLingual~\cite{deng2023multilingual}  & Black            & User               & Long-tail: low-resource  & High        & $\checkmark$       & Harmful Content \\
LRL~\cite{yong2023low}  & Black            & User           & Long-tail: low-resource  &  High       & $\checkmark$       & Harmful content \\
CodeChameleon~\cite{lv2024codechameleon} & Black            & User                  & Long-tail: encrypts     &  High         &  $\checkmark$    & Harmful Content             \\
ReNeLLM~\cite{ding2023wolf}      & Black      & User      & Prompt rewriting \& scenario nesting            &  High     &  $\checkmark$               & Harmful Content                 \\
PAIR~\cite{chao2023jailbreaking}        & Black            & System             &  Automatic Iterative Refinement       & High        &  $\checkmark$    &  Harmful content          \\
GPTFUZZER~\cite{yu2023gptfuzzer}   & Black            &  User                 & Fuzzing                 &  Low             & $\checkmark$    &  Harmful content            \\
TAP~\cite{mehrotra2023tree}          & Black            &  System                 & Tree-of-thought reasoning        & Medium         & $\checkmark$  &  Harmful content                \\
Mosaic Prompts~\cite{glukhov2023llm}      & Black            & User    & Semantic censorship                        & High            & $\times$   & Impermissible content   \\
EasyJailbreak~\cite{zhou2024easyjailbreak} & Black            &  System+User               &  Unified framework for 12 jailbreaks    &  -    & $\checkmark$ & Jailbreak attack evaluation         \\
PROMPTINJECT~\cite{perez2022ignore}  & Black       & User    & Mask-based iterative strategy     & Low     &  $\times$          &   Goal hijacking \& prompt leaking              \\
IPI~\cite{greshake2023not}         & Black                 &  System      & Indirect prompt injection                       & High           &  $\checkmark$   & Cyber threats like theft of data and denial of service etc.            \\
HOUYI~\cite{liu2023prompt}       & Black  & User               &  SQL injection \& XSS attacks                  & Low            &   $\times$            & Prompt abuse \& prompt leak \\
GA~\cite{lapid2023open}& Black    & User                  &  genetic algorithm        &  Low      & $\times$     & Harmful content 
\\
GUARD~\cite{jin2024guard} & Black   & System+User                &  Role-playing LLMs                   & High             & $\times$  & Harmful content
\\
CIA~\cite{jiang2023prompt}& Black      & User                  & Combination of multiple instructions                      & Medium           & $\checkmark$  & Harmful content  
\\

\midrule
Pelrine et al. \cite{pelrine2023exploiting}~$\textbf{*}$& Grey                 &  User                  & fine-tuning                   &  Low & $\checkmark$               &  Misinformation \& Privacy Leakage 
\\
Zhang et al. \cite{zhan2023removing} & Grey                 &  User                  & fine-tuning                        &  Low & $\checkmark$               &  Harmful content 
\\
Safety-tuned \cite{bianchi2023safety} & Grey                 &  User                  & fine-tuning                        &   Low               &  $\times$ & Harmful content
\\
Janus inference \cite{chen2023janus}& Grey                 &    System &  Fine-tuning             &    Low                     &    $\times$             & Privacy Leakage 
\\
Qi et al.\cite{qi2024finetuning} & Grey                 &     User               & fine-tuning                         &    Low            &  $\times$ & Harmful content
\\
Pelrine et al. \cite{pelrine2023exploiting}& Grey                 &     User                  & poisoning knowledge retrieval                        &  Medium & $\checkmark$               &  Harmful content \& fairness  
\\
PoisonedRAG\cite{zou2024poisonedrag}& Grey                 &  System                  & poisoning knowledge retrieval                        & Low  & $\checkmark$               &  Hallucination
\\
AutoPoison~\cite{shu2024exploitability}& Grey                 &    System                &   Content Injection                      & Low & $\times$                &   Triggered response
\\
LoFT~\cite{attacksloft}& Grey                 &   System                 &    fine-tuning                     &          Low       &  $\checkmark$ & Harmful content
\\
BadGPT~\cite{shi2023badgpt}& Grey                 &  System                  &   Backdoor Attack                      &   Medium              & $\times$ &  Harmful content
\\
ICLAttack~\cite{zhao2024universal}& Grey                 &  System                  &   Backdoor Attack                      &   Low              & $\times$ &  Triggered response
\\
ActivationAttack~\cite{wang2023backdoor}& Grey                 &  System                  &   Activation Steering                      &   Low              & $\times$ &  Harmful \& Biased content
\\
\bottomrule[1.5pt]
\end{tabular}
}
\flushleft * Claim to Jailbreak the Guardrails
\end{table*}

\subsection{White-box Jailbreaks}\label{sec:white}
A white-box attack normally refers to a scenario in which the attacker has full access to the internal details of the model.
Since even LLMs with guardrails can not fully protect against adversarial attacks, we introduce some techniques for attacking LLMs (with or without guardrails) under the white-box setting.
Notably, most white-box attacks can be applied to the \textit{black-box} scenario by employing their transferability on a white-box surrogate model.

\subsubsection{Learning-based Methods}

The \textbf{Greedy Coordinate Gradient (GCG)}~\cite{zou2023universal} method was designed to search a specific sequence of characters (an \textit{adversarial suffix}).
When the adversarial suffix is attached to different queries, it misleads the LLM to generate a response with harmful content. 
This approach integrates greedy search and gradient-based methods for discrete optimization to manipulate the model's outputs. 
It aims to optimize the likelihood that the model will generate an affirmative response, e.g., ``Sure, this is...''. To improve the computational efficiency of GCG, ~\cite{geisler2024attacking} revisited Projected Gradient Descent (\textbf{PGD}) for LLMs, which has been widely used for generating adversarial perturbations in other domains, by controlling the error introduced by the continuous relaxation for the input prompt, it can fool LLMs with the similar attack performance but with up to one order of magnitude faster. Previous techniques in traditional NLP, like Gradient-based Distributional Attack (\textbf{GBDA})~\cite{guo2021gradient}, can also be used to search adversarial suffixes.
Specifically, it also applies continuous relaxation usingGumbel-Softmax~\cite{jang2016categorical},
which allows for the manipulation of text inputs in a gradient-guided manner, maintaining the textual data's discrete nature while optimizing the adversarial objective.
However, it fails to obtain high jailbreaking performance on the aligned LLMs.

On the other hand, the adversarial suffixes produced by GCG are mostly some garbled characters that are easily detectable by simple perplexity filter~\cite{jain2023baseline}. \textbf{AutoDAN-Zhu}~\cite{zhu2023autodan} design a double-loop optimization method upon GCG to generate more stealthy jailbreak prompts.
In addition, it also demonstrates the ability and interpretability to solve other new tasks like prompt leaking.
Furthermore, \textbf{COLD-Attack}~\cite{guo2024cold} adopts a new jailbreak, 
to automate the search of adversarial LLM attacks under a variety of restrictions such as fluency, stealthiness, sentiment, and left-right-coherence.
It performs efficient gradient-based sampling in the continuous logit space and relies on a guided decoding process to translate the continuous logit sequences back into discrete text. 

Although the adversarial suffix can lead the base LLM to generate harmful responses, the LLM models with guardrails can easily detect it. Mangaokar et.al.~\cite{mangaokar2024prp} proposed a new attack strategy, named \textbf{PRP}, for attacking LLM models with guardrails mainly. 
Specifically, it leverages a two-step prefix-based attack, including universal adversarial prefix construction and prefix propagation to the response. Inserting the universal prefix into the response can elicit the guardrail for outputting the harmful content.
After the universal prefix generation, the corresponding propagation prefix can be created through a few in-context templates. Such in-context learning enables the LLM to initially output the pre-computed or desired adversarial prefix, eventually making PRP jailbreak the LLM models with guardrails. 
Subsequently, \textbf{AutoDAN-Liu~\cite{liu2023autodan}}proposed to generate stealthy jailbreak prompts automatically, it utilizes a hierarchical genetic algorithm to bypass the ethical guidelines and safety measures of LLMs. 
This method is grounded in optimization techniques inspired by natural selection. It iteratively refines generations of prompts to circumvent built-in safeguards effectively. Through this evolutionary process, AutoDAN-Liu generates stealthy prompts that subtly avoid triggering the model's protective mechanisms. 



\subsubsection{LLM Generation Manipulatation}

On the other hand, except for jailbreaking via some learning-based method, ProMan~\cite{zhang2023safety} was proposed to directly manipulate the generation process of those open-source LLMs and enforce the LLMs to generate specific tokens at specific positions, therefore cheating the LLMs to generate 
undesired response, including harmful or sensitive
information or even private data.


Although the current literature only describes limited white-box jailbreaks, it is still possible to bypass the guardrails if full access is provided.
For example, suppose the attacker knows the guardrail used for the targeted LLM is Llama Guard, and the adversary has full access to the fine-tuned Guard model. 
In that case, the previous white-box attacks using optimization can be further extended: the optimization space will be further narrowed by adding an extra constraint, i.e., the adversarial input and/or the resulting response of LLMs are supposed to be evaded by the guardrail.
In other words, a successful jailbreak should satisfy the safety conditions of the targeted LLM and the Guard models.




\subsection{Black-box Jailbreaks}\label{sec:black}


Unlike white-box attacks, which necessitate access to model weights and tokenizers, black-box attacks operate under the premise that adversaries lack knowledge of the LLM's internal architecture or parameters. Therefore, they are more common. 
In this subsection, jailbreak attacks conducted within a black-box setting are classified into four categories: i) \textit{manually designed jailbreaks}, ii) \textit{attacks exploiting long-tail distribution}, iii) \textit{optimization-based methods for generating jailbreaks}, and iv) \textit{unified framework for jailbreaking}.


\subsubsection{Delicately Designed Jailbreaks}

The phenomenon of jailbreak attacks against state-of-the-art large LLMs was investigated in \textbf{JailBroken} \cite{wei2024jailbroken}, explicitly focusing on models such as GPT-4, GPT-3.5 Turbo, and Claude v1.3. This work identifies two primary reasons for the successful attack: competing training objectives and instruction tuning objectives. The authors propose these two failure modes as guiding principles for designing new jailbreak attacks. Using carefully engineered objectionable prompts, they empirically evaluate these attacks against the aforementioned safety-trained LLM models. The results indicate a high success rate regarding a large number of the attacks.

In this line, due to the constant evolution of ethical and legal constraints embedded within LLM safeguards, jailbreak attempts employing direct instructions \cite{wei2024jailbroken, chao2023jailbreaking} are typically easily identified and rejected. It is motivated by the Milgram shock experiment \cite{milgram1963behavioral, wenglinsky1975obedience} and its adaptation to LLMs, which follow authoritative commands to produce harmful content. A prompt-based jailbreak method, called \textbf{DeepInception}\cite{li2023deepinception}, was devised for conducting a black-box attack on LLMs. This involves injecting an inception mechanism into a LLM and effectively hypnotizing it to act as a jailbreaker. DeepInception explicitly constructs a nested scene to serve as the inception for guiding the behavior of the LLM. This nested scene facilitates an adaptive approach to circumvent safety constraints in a normal scenario, opening up the potential for subsequent jailbreaks. Specifically, DeepInception utilizes the personification capability of LLMs, along with their tendency to follow instructions, to generate diverse scenes or characters.  \cite{shen2023anything} furthered this by creating prompts that encourage ChatGPT to act as a \textbf{DAN} (``Do Anything now"). As implied by their designation, LLMs are now capable of boundless functions. They are no longer bound by the customary rules that govern AI systems.

While existing attack methods are typically applied to new conversations devoid of context, the potential of In-Context Learning (ICL) was delved into the influence of the alignment ability of LLMs. Leveraging these insights, the study introduces the \textbf{In-Context Attack (ICA)} \cite{wei2023jailbreak}. ICA is tailored to construct malicious contexts to direct models to produce harmful outputs. The efficacy of in-context demonstrations in aligning LLMs is demonstrated, and implementing these methods is straightforward. Additionally, Deng et.al.~\cite{deng2023attack} proposed a semi-automatic attack framework named Semi-Automatic Attack Prompt (\textbf{SAP}), it combines manual and automatic methods to generate prompts to mislead LLMs to output harmful content. 
Specifically, they manually construct high-quality prompts as an initial prompt set and then iteratively update them through in-context learning with LLMs.
Through this red-teaming attack, extensive high-quality attack prompts can be efficiently generated. Liu et.al.~\cite{liu2024making} proposed a novel universal jailbreak approach named \textbf{DRA} (Disguise and Reconstruction Attack. 
This method involves concealing harmful instructions via disguise, prompting the model to uncover and reconstruct the original harmful instruction within its generated output, thus navigating around traditional security measures.
In this way, the harmful input can be disguised from the input filter, which then guides the target to reconstruct the attack to obtain the desired response from the adversary.

\subsubsection{Exploiting Long-tail Distribution}

Jailbreaks relying on long-tail distributed encoding convert the original query into rare or unique data formats such as ciphers \cite{yuan2023gpt}, low-resource languages \cite{deng2023multilingual}, and personalized encryption methods \cite{lv2024codechameleon}. The safety vulnerability of LLMs when user queries are encrypted was investigated in \textbf{CipherChat }\cite{yuan2023gpt}. The framework involves encoding malicious unsafe text using LLMs and assessing the safety of the decoded responses. CipherChat is designed with three key elements in its system prompt to ensure effective communication through ciphers: (1) behavior assignment, (2) cipher teaching, and (3) enciphered unsafe demonstrations. It enables users to interact with LLMs using cipher prompts, system role descriptions, and few-shot enciphered demonstrations. Furthermore, the authors introduce SelfCipher, which utilizes a hidden cipher embedded within LLMs to circumvent safety features more efficiently than existing ciphers.

Afterward, despite the widespread use of English globally, there is growing concern that the safety of LLMs is predominantly assessed in English alone. However, \textbf{MultiLingual} \cite{deng2023multilingual} takes a significant stride forward by investigating the safety levels of LLMs across various languages, including those with limited linguistic resources. This research delves into the vulnerabilities of LLMs from two perspectives: unintentional and intentional scenarios. In the unintentional scenario, queries translated into non-English languages unexpectedly expose users to unsafe content. Conversely, the intentional scenario involves using translated multilingual ``jailbreak'' prompts. 
Similarly, Low Resource Languages-Combined Attacks~\cite{yong2023low} (\textbf{LRL}) also underlines the cross-lingual vulnerability of GPT-4. 
By translating unsafe English prompts into less commonly used languages, they successfully circumvent protective measures to elicit harmful responses. \cite{kang2023exploiting} show that instruction-following language models using the \textsc{text-davinci-003} prompt could potentially be employed to produce malicious content.

A hypothesis regarding LLMs' safety mechanisms was proposed in subsequent research, suggesting that LLMs first detect intent before generating responses. Building on this hypothesis, a framework was introduced known as \textbf{CodeChameleon}, which encrypts and decrypts queries into a format challenging for LLMs to detect \cite{lv2024codechameleon}. Four distinct encryption functions are employed during the encryption stage based on reverse order, word length, odd and even positions, and binary tree structure. Subsequently, the decryption functions are incorporated into the instructions as code blocks. During inference, these decryption functions assist LLMs in understanding the encrypted content. Extensive testing demonstrates that CodeChameleon effectively circumvents LLMs' intent recognition.


\subsubsection{Optimization-based Approaches}

In contrast to conventional adversarial examples, such jailbreaks are usually created through human ingenuity, strategically devising situations that naturally mislead the models \cite{wei2024jailbroken}, rather than relying on automated techniques. Consequently, crafting them demands considerable manual labor. The adversarial prompts generated \textbf{Greedy Coordinate Gradient (GCG)} \cite{zou2023universal} exhibit a high degree of universality and transferability, particularly to other fully black-box models. 

To avoid limitations regarding intricate manual design \cite{wei2024jailbroken, yuan2023gpt} and require optimization on other white-box models, compromising generalization or efficiency \cite{zou2023universal}, a method known as \textbf{ReNeLLM} was introduced \cite{ding2023wolf}. ReNeLLM is an automatic jailbreak prompt generation framework, which generalizes jailbreak prompt attacks into two aspects: (1) Prompt Rewriting and (2) Scenario Nesting. 

Following this, \textbf{Prompt Automatic Iterative Refinement (PAIR)} proposed an automated red teaming method for jailbreaking LLMs \cite{chao2023jailbreaking}, which represents a significant improvement of over ten thousand times compared to existing attacks, such as jailbreaks identified through Greedy Coordinate Gradient (GCG) \cite{zou2023universal}. The authors aim to find a balance between prompt-level attacks \cite{dinan2019build}, which are labor-intensive but scalable, and token-level attacks \cite{maus2023black}, which are uninterpretable and inefficient in terms of queries. PAIR devised a protocol leveraging a language model to craft prompt-level attacks that are both semantic and human-interpretable. This involves an automated system where the attacker language model learns from prior prompts and responses to refine based on a judge score and generate new prompts. Through in-context learning, PAIR enabled the language model to enhance the quality of generated candidate queries autonomously.

Drawing inspiration from AFL fuzzing, \textbf{GPTFUZZER }was introduced, a black-box jailbreak fuzzing framework to autonomously generate jailbreak prompts \cite{yu2023gptfuzzer}. GPTFUZZER aims to combine the efficacy of human-written prompts with the scalability and flexibility of automated systems to bolster the assessment of vulnerabilities in LLMs. The framework is built upon a seed selection strategy, mutate operators, and a judgment model. By harnessing these elements, GPTFUZZER can systematically detect and exploit vulnerabilities in LLMs.

Building upon prior automated methodologies, \textbf{Tree of Attacks with Pruning (TAP)} introduced a novel approach for generating jailbreaks \cite{mehrotra2023tree}. TAP leverages an LLM to iteratively refine candidate prompts using tree-of-thought reasoning until a successful jailbreaking prompt is generated. The framework involves three key components: an attacker LLM tasked with generating jailbreaking prompts using tree-of-thought reasoning, an evaluator responsible for assessing the generated prompts and determining the success of the jailbreaking attempt, and a target LLM that serves as the subject of the jailbreaking endeavor. Lapid et.al.~\cite{lapid2023open} employed the genetic algorithm (\textbf{GA}) for generating a universal adversarial suffix under the black-box setting.
Instead of maximizing the targeted token likelihOOD in GCG, they proposed using random subset sampling for fitness approximation by minimizing the cosine similarity between benign input embedding and adversarial input embedding. 
Experiments illustrate high transferability across different LLMs. Furthermore, Jin et.al.~\cite{jin2024guard} proposed a role-playing system named Guideline Upholding through Adaptive Role-play Diagnostics (\textbf{GUARD}), which allocates four distinct roles to user LLMs to collaborate on new jailbreaks.
By collecting some existing jailbreak prompts into a knowledge graph and using Chain-of-Thought to align with the specific functions and objectives for each role, they can generate a higher jailbreak success rate and a lower perplexity score than GCG~\cite{zou2023universal} and AutoDAN~\cite{zhu2023autodan}.

\subsubsection{Unified Framework for Jailbreaking}

A recent development, EasyJailbreak \cite{zhou2024easyjailbreak}, presents a comprehensive framework to evaluate jailbreak attacks on LLMs. This framework integrates four pivotal components: Selector, Mutator, Constraint, and Evaluator. This approach allows researchers to concentrate on crafting unique components, thus minimizing the effort required for development. Moreover, it demonstrates broad model compatibility, accommodating various models, including open-source alternatives like LlaMA2 and proprietary ones like GPT-4.


\subsubsection{Prompt Injection for Desired Responses}

Prompt Injection in LLMs involves the malicious alteration of input provided to the model, commonly achieved by substituting original instructions with carefully crafted user input \cite{shayegani2023survey}. This manipulation occurs within the framework of supplying prompts to the LLM, guiding its responses or behaviors. Prompt injection attacks present a significant cybersecurity risk as they can result in creating unauthorized content, circumventing content moderation protocols, exposing sensitive data, or even facilitating the dissemination of malicious code or malware. This vulnerability is particularly prominent in LLMs that employ prompt-based learning approaches, rendering them susceptible to exploitation by malicious attackers. Given the significant role of prompt in shaping LLM output, prompt injection manipulation can have widespread implications for attacking LLMs.

Since most LLMs, such as ChatGPT, are closed-source platforms, much of the research centers on utilizing prompt engineering techniques to induce ChatGPT to generate inappropriate content.
The framework known as \textbf{PROMPTINJECT} \cite{perez2022ignore} was proposed as a straightforward alignment mechanism for generating iterative adversarial prompts through masks. This approach involves assembling prompts to facilitate a quantitative assessment of the robustness of LLMs against adversarial prompt attacks. The study focuses primarily on evaluating the susceptibility of GPT3 to such attacks, accomplished through simplistic handcrafted inputs. The analysis concentrates on two types of attacks: goal hijacking and prompt leaking. Goal hijacking involves introducing a malicious string, termed a rogue string, designed to divert the model into generating a particular sequence of characters. Conversely, prompt leaking pertains to the possibility of revealing a private value embedded within a confidential prompt, which should not be disclosed externally under any circumstances. Following this, the concept of \textbf{Indirect Prompt Injection (IPI)} \cite{greshake2023not} was introduced, referring to an uninvestigated attack vector where retrieved prompts can function as ``arbitrary code'', thus compromising LLM-integrated applications. The authors demonstrate these attacks against real-world systems like Bing Chat, code-completion engines, and GPT-4. 

Inspired by traditional injection attacks, a novel black-box prompt injection attack technique called \textbf{HOUYI} \cite{liu2023prompt} was introduced. HOUYI comprises three essential components: a preconstructed prompt, an injection prompt, and a malicious question, each tailored to achieve the adversary’s goals. Two significant exploit scenarios were identified: prompt abuse and prompt leak. The application of HOUYI to a sample of 36 real-world LLM-integrated applications revealed that 31 of these applications are vulnerable to prompt injection. \cite{glukhov2023llm} raised a concept of semantic censorship, which falls into the category of guardrail using
a universal algorithm to determine whether the content generated by an LLM is permissible based on semantic content alone.
Accordingly, they proposed a novel attack named \textbf{Mosaic Prompts}; it leverages the ability of a user to query an LLM multiple times in independent contexts to construct impermissible outputs from a set of permissible ones.
This indicates a significant limitation of output censorship, as it cannot provide safety or security guarantees without imposing severe restrictions on model usefulness. Moreover, Compositional Instruction Attack (\textbf{CIA})~\cite{jiang2023prompt} capitalizes on LLMs' failure to detect underlying harmful intents when instructions are composed of multiple elements, thus revealing significant vulnerabilities in LLM security mechanisms.
They outline two specific strategies, Talking-CIA (T-CIA) and Writing-CIA (W-CIA), developed to automate the generation of these deceptive instructions. 
T-CIA leverages psychological principles to align the model's response persona with the harmful intent, bypassing LLMs' ethical constraints. 
Conversely, W-CIA disguises harmful prompts as creative writing tasks, exploiting LLMs' lack of judgment on fictional content to elicit dangerous outputs.

\subsection{Gray-box Jailbreaks}\label{sec:grey}

In \cite{pelrine2023exploiting}, the authors highlight that beyond the white-box approach, which involves full access to a model's parameters, and the more limited black-box method, there's also `grey-box' access. This middle ground could be crucial in uncovering additional vulnerabilities in the safeguard systems of LLMs. This section will present studies on `grey-box' attack methods, encompassing strategies like fine-tuning, retrieval-augmented generation, and backdoor attacks.
\subsubsection{Fine-tuning Attacks}
Fine-tuning technology enables users to customize pre-trained LLMs effectively. However, when these fine-tuning privileges are extended to end-users, the existing guardrails may not be sufficient to prevent harmful behaviors. The attacks by fine-tuning the LLMs can also be called 'grey-box' attacks. \cite{zhan2023removing} suggested that fine-tuning could mitigate Reinforcement Learning with Human Feedback (RLHF) safeguards, commonly employed in LLMs to minimize harmful outputs. Their research revealed that even ChatGPT 4 could have its protections removed by fine-tuning. Through experiments, they demonstrated a success rate of 95\% in generating harmful responses from ChatGPT 4, using just 340 examples for fine-tuning. The experiments from \cite{pelrine2023exploiting} also indicate that fine-tuning a model with only 15 harmful or 100 benign examples can compromise the safeguards of GPT-4, leading to a variety of harmful outputs. \cite{bianchi2023safety} also examines the potential safety risks associated with excessive instruction tuning in LLM, illustrating that models excessively tailored to specific instructions may still generate harmful content. To counteract these risks, the researchers suggest developing a safety-focused tuning dataset to balance the dual objectives of maintaining model performance while enhancing safety measures.
Furthermore, the research by \cite{chen2023janus} highlights the risks of fine-tuning language models using small datasets containing personally identifiable information (PII). Initially, it focuses on a simple approach where a language model is fine-tuned with a small dataset rich in text-based PII, which results in the model being more likely to divulge PII upon prompting. Then, the researchers introduced the "Janus" methodology, which centers around defining a PII recovery task followed by few-shot fine-tuning. Experimental findings demonstrate that fine-tuning GPT-3.5 with just 10 PII examples markedly increases the model's ability to expose PII. \cite{qi2024finetuning} found that additional training of the model can compromise the effectiveness of established guardrails. They bypass the GPT-3.5 Turbo's safety guardrails by fine-tuning it with justten0 specific examples and successfully make the model entirely susceptible to harmful instructions.  

\subsubsection{Retrieval-Augmented Generation (RAG)}
RAG for LLMs aims to improve the response of LLMs by incorporating external datasets during inference. It integrates context and up-to-date or relevant information in the prompt to enhance the LLM's performance. \cite{pelrine2023exploiting} finds that employing the prompt injection techniques suggested by \cite{perez2022ignore} indicated that polluting the external dataset by injecting a malicious instruction could successfully invalidate ChatGPT 4's safety protection. They also demonstrated that if biased system messages accompany the upload of factual data, it can bias the responses of ChatGPT. \cite{zou2024poisonedrag} also proposed to inject toxic texts into the knowledge database to compromise LLMs. They developed these poisoned texts by forming them to solve an optimization problem aimed at generating a target response chosen by the attacker. Their experiments showed that by injecting just five tainted texts tailored to a specific question, they were able to attain a 90\% attack success rate.

\subsubsection{Backdoor Attack}
The backdoor attack on the neural language process task is to manipulate the model to produce specific outputs when triggered ~\cite{cai2022badprompt}.
 It typically occurs during the pre-training and adaptation tuning, where the backdoor trigger gets embedded \cite{chen2021badnl}. These manipulations should maintain the model's performance and evade detection by human inspection. The backdoor is triggered exclusively when input prompts to LLMs include the embedded trigger, causing the compromised LLMs to act maliciously as intended by the attacker. \cite{shu2024exploitability} propose Auto Poinson to incorporate training examples that reference the desired target content into the system, triggering similar behaviors in downstream models. \cite{attacksloft} introduces LoFT (Local Proxy Fine-tuning) to fine-tuning smaller, local proxy models to develop attacks that are more likely to transfer successfully to larger, more complex LLMs. This technique leverages the target LLMs to produce prompts closely aligned with harmful queries, effectively gathering prompts from a localized vicinity around these queries. A set of parameters in the proxy LLM is then fine-tuned, guided by the responses of the target LLM to these analogous prompts. Ultimately, this fine-tuned proxy model is deployed to attack the target LLMs. The study demonstrates that this method improves the transferability of attacks. \cite{shi2023badgpt} proposed the BadGPT, a backdoor attack targeting RL fine-tuning in language models. It injects a backdoor trigger into the reward model during the fine-tuning stage, allowing for compromising the fine-tuned language model.  \cite{zhao2024universal} then proposed ICLAttack, which fine-tunes models by targeting in-context learning for backdoor attacks. This method focuses on two prompt-level strategies: introducing compromised examples within the prompt's demonstration set and modifying the prompts. This technique operates at the prompt level, eliminating the necessity to train new LLMs altogether. On the other hand, Wang et.al.\cite{wang2023backdoor} pointed out that poisoning the training dataset or introducing harmful prompts affects the adaptability of the attacks, rendering them more prominent and more accessible to identify. They propose using activation steering without optimization to target four key aspects of LLMs: truthfulness, toxicity, bias, and harmfulness.







\subsection{Techniques for Strengthening LLMs}
This section discusses techniques that may help construct more powerful defenses for guardrails or more robust LLMs.

\subsubsection{Detection-based Methods: Guardrail Enhancement}

 To detect the harmful information in the user's input, \textbf{PPL}~\cite{alon2023detecting} calculates the perplexity of a provided input to decide whether a user's request should be accepted or rejected. \textbf{SmoothLLM}~\cite{robey2023smoothllm} borrowed the idea of randomized smoothing literature~\cite{cohen2019certified}, it
randomly alters multiple versions of a given input and then combines the respective predictions to identify adversarial inputs. Some researchers have explored how In-Context Learning (ICL) can impact the alignment capabilities of LLMs. \textbf{In-Context Defense (ICD)} \cite{wei2023jailbreak} method is designed to bolster model resilience by demonstrations of rejecting to answer harmful prompts via in-context demonstration.

To defend LLM attacks, \textbf{LLM SELF DEFENSE}~\cite{helbling2023llm} was proposed first. Specifically, by incorporating the generated content into a pre-defined prompt and using another instance, LLM, to analyze the text, it constructs an extra guardrail filter for preventing harmful content. Furthermore, Cao et.al.~\cite{cao2023defending} proposed Robustly Aligned LLM (\textbf{RA-LLM}) to defend against potential alignment-breaking attacks.
Unlike the previous alignment check, which uses the alignment check function to decide whether to reject the response, the proposed Robust Alignment Check Function adds several extra random droppings on the request. It usually checks whether the corresponding response can still pass the alignment check function AC. Then, Chen et.al.~\cite{chen2023jailbreaker} designed a moving target defense (\textbf{MTD}) to enhance the LLM system.
Compared to previous guardrail methods that decide whether the input/output is safe, MTD calculates a composite score for each response by combining its \textit{quality} and \textit{toxicity} metrics.
It employs randomization to select a response that qualifies both response metrics, eventually providing a solid moving target defense for the LLMs.

\subsubsection{Mitigation-based Methods: Affirmative Response Generation}

As shown in.~\cite{jain2023baseline}, besides perplexity filtering, input preprocessing like \textbf{Retokenization} and \textbf{Paraphrase} can also successfully compromise the effectiveness of some attacks like GCG~\cite{zou2023universal}.
However, adversarial training, though once favored for safeguarding image classifiers, faces diminished appeal for  LLMs due to the prohibitive expenses associated with both model pre-training and the creation of adversarial attacks, rendering large-scale adversarial training impractical.
Finding a good approximation
for robust optimization objectives that allow for successful adversarial training remains an open challenge. Further, Li et.al.~\cite{li2023rain} introduced a novel inference method,
Rewindable Auto-regressive INference (\textbf{RAIN}) enables pre-trained LLMs to assess their own outputs and leverage the assessment outcomes to inform and steer the backtracking and generating content to enhance AI safety.
Contrary to Reinforcement Learning from Human Feedback (RLHF), RAIN dispenses with the requirement for extra model upkeep and bypasses the accumulation of gradient data and computational graphs. Still, it must pay the extra but acceptable cost of the auto-regressive inference. Additionally, Zhang et.al.~\cite{zhang2023defending} proposed to integrate goal prioritization (\textbf{GP}) at both training and inference stages.
It analyzes the reason behind successful jailbreaking: the conflict between two goals: helpfulness (providing helpful responses to user queries) and safety (providing harmless and safe responses to user queries.
The jailbreak attack success rate can be notably decreased by plugging in the goal prioritization for these two properties into the inference alone or with training.

Further, \textbf{Self-Reminder}~\cite{xie2023defending} suggests that adding self-reminder prompts can be an effective defense.
They speculate that initiating ChatGPT with a `system mode' prompt at the most external level to remind it of its role as a responsible AI tool could reduce its vulnerability to being malevolently steered by user inputs at a deeper level.
Therefore, by concatenating an extra system prompt after the user’s query that reminds the LLMs to respond responsibly, the experimental results showed that self-reminders significantly reduce the success rate of jailbreak attacks. Then, Ge et.al.~\cite{ge2023mart} proposed a multi-round automatic red-teaming framework \textbf{MART} to improve the scalability of safety alignment. 
Two players, i.e., an adversarial LLM and a target LLM, iteratively interplay with each other.
The adversarial LLM aims to generate challenging prompts that provoke unsafe responses from the target LLM. Concurrently, the target LLM is refined with data that aligns with safety standards based on these adversarial prompts.
Through several rounds of red-teaming, the enhanced target LLM continues to bolster its defenses through safety-specific fine-tuning. Further, Zhou et.al.~\cite{zhou2024robust} proposed the first adversarial objective aimed at protecting language models 
from jailbreaking attacks, along with a novel algorithm, Robust Prompt Optimization (\textbf{RPO}). 
This strategy employs gradient-based token optimization (similar to GCG) to ensure the generation of harmless outputs.
RPO represents the initial approach in jailbreaking defense (like adversarial training in vision) that enhances robustness comprehensively and effectively and at only a minor cost to normal use.
\textbf{SafeDecoding}~\cite{xu2024safedecoding} found that despite the likelihood of tokens signifying harmful content being higher than those for harmless responses, safety disclaimers continue to emerge among the highest-ranking tokens when sorted by probability in descending order.
Thus, in the training phase, the model will be fine-tuned with a few safety measures, and then, during the inference, SafeDecoding further constructs the new token distribution.
The crafted probability distribution reduces the chances of tokens that resonate with the attacker's objectives and enhances the probabilities of tokens that align with human values.





\section{Discussions: A Complete Guardrail}\label{sec:5}

Based on the discussions about tackling individual requirements in Sections~\ref{sec:3} and \ref{sec:4}, this section advocates building a guardrail by systematically considering multiple requirements.
We discuss four topics:  conflicting requirements (Section~\ref{sec:conflicting}),  multidisciplinary approach  (Section~\ref{sec:sociotechnical}),  implementation strategy (Section~\ref{sec:neuralsymbolic}), rigorous engineering process  (Section~\ref{sec:sdlc}), and safeguards for LLM Agents (Section~\ref{sec:5.5}).  

\subsection{Conflicting Requirements}
\label{sec:conflicting}

This section discusses the tension between safety and intelligence as an example of the conflicting requirements. Conflicting requirements are typical, including, e.g., fairness and privacy \cite{xiang2022being}, privacy and robustness \cite{10.1145/3319535.3354211}, and robustness and fairness \cite{Bassi2024}. Integrating guardrails with LLMs may lead to a discernible conservative shift in the generation of responses to open-ended text-generation questions \cite{rottger2023xstest}. 
The shift has been witnessed in ChatGPT over time. \cite{chen2023chatgpt} documented a notable change in ChatGPT's performance between March and June 2023. Specifically, when responding to sensitive queries, the model's character count decreased significantly, plummeting from an excess of 600 characters to approximately 140. Additionally, in the context of opinion-based questions and answers surveys, the model is more inclined to abstain from responding.

Given the brevity and conservativeness of responses generated by ChatGPT, the following question arises: How can exploratory depth be maintained in responses, particularly for open-ended test generation tasks? Furthermore, does the application of guardrails constrain ChatGPT's capacity to deliver more intuitive responses? On the other hand, \cite{NarayananKapoor2023GPT4}  critically examined this paper and emphasized the difference between an LLM's capabilities and its behavior. Although capabilities typically remain constant, behavior can alter due to fine-tuning, which can be interpreted as the ``uncertainty'' challenges in LLMs. They suggest that GPT-4's performance changes are likely linked more to evaluation data and fine-tuning methods rather than a decline in its fundamental abilities. They also acknowledge that such behavioral drift challenges the development of reliable chatbot products. The adoption of guardrails has also led to the model adopting a more concise communication approach, offering fewer details and electing non-response in specific queries. The decision ``to do or not to do'' can be challenging when designing the guardrail. While the most straightforward approach is to decline an answer to any sensitive questions, is it the most intelligent one? That is, \emph{we need to determine if the application of guardrail always has a positive impact on LLMs that is within our expectation}. 

\textbf{Our Perspective}
 Prior research suggested incorporating a creativity assessment mechanism into the guardrail development for LLMs. 
 To measure the creativity capability of LLMs, \cite{chakrabarty2023art} employed the Consensual Assessment Technique \cite{amabile1982social}, a well-regarded approach in creativity evaluation, focusing on several key aspects: fluency, flexibility, originality, and elaboration, which collectively contribute to a comprehensive understanding of the LLMs' creative output in storytelling. \cite{NarayananKapoor2023GPT4} showed that although some LLMs may demonstrate adeptness in specific aspects of creativity, there is a significant gap between their capabilities and human expertise when evaluated comprehensively. 
\subsection{Multidisciplinary Approach} \label{sec:sociotechnical}

While current LLM guardrails include mechanisms to detect harmful content, they still risk generating biased or misleading responses. It is reasonable to expect future guardrails to integrate harm detections and {other mechanisms to deal with, e.g., ethics, fairness, and creativity}. In the introduction, we have provided three categories of requirements to be considered for a guardrail. Moreover, LLMs may not be universally effective across all domains, and it has been a trend to consider domain-specific LLMs \cite{Soumen2023}. In domain-specific scenarios, specialized rules may conflict with the general principles. For instance, in crime prevention, the use of certain terminologies that are generally perceived as harmful, such as `guns' or `crime,' is predominant and should not be precluded. To this end, the concrete requirements for guardrails will differ across different LLMs, and research is needed to \emph{scientifically} determine requirements.
The above challenges (multiple categories, domain-specific, and potentially conflicting requirements) are compounded by the fact that many requirements, such as fairness and toxicity, are hard to define precisely, especially without a concrete context. The existing methods, such as the popular one that sets a threshold on predictive toxicity level \cite{perez2022red}, do not have \emph{valid justification and assurance}. 

\textbf{Our Perspective}
Developing LLMs ethically involves adhering to fairness, accountability, and transparency. These principles ensure that LLMs do not perpetuate biases or cause unintended harm. The works by 
e.g., \cite{sun2023aligning} and 
\cite{DBLP:journals/corr/abs-2312-11779} provide insights into how these principles can be operationalized in the context of LLMs. Establishing community standards is vital for the responsible development of LLMs. These standards, derived from a consensus among stakeholders, including developers, users, and those impacted by AI, can guide LLMs' ethical development and deployment. They ensure that LLMs are aligned with societal values and ethical norms, as discussed in broader AI ethics literature \cite{AF2023LLM}.
Moreover, the ethical development of LLMs is not a one-time effort but requires ongoing evaluation and refinement. These tasks involve regular assessment of LLMs outputs, updating models to reflect changing societal norms, and incorporating feedback from diverse user groups to ensure that LLMs remain fair and unbiased.

Socio-technical theory \cite{trist1957studies}, in which both `social' and `technical' aspects are brought together and treated as interdependent parts of a complex system, have been promoted \cite{10266809, DBLP:journals/corr/abs-2006-09663} for machine learning to deal with properties related to human and societal values, including e.g., fairness \cite{https://doi.org/10.1111/isj.12370}, biases \cite{schwartz2022}, and ethics \cite{mbiazi2023survey}. To manage the complexity, the whole system approach \cite{crabtree2011chronic}, which promotes an ongoing and dynamic way of working and enables local stakeholders to 
come together for an integrated solution,  has been successfully working on healthcare systems \cite{brand2017whole}. We believe a multi-disciplinary group of experts will work out and rightly justify and validate the concrete requirements for a specific context by applying the socio-technical theory and the whole system approach.

\subsection{Neural-Symbolic Approach for Implementation}\label{sec:neuralsymbolic}

Existing guardrail frameworks such as those introduced in Section~\ref{sec:3} employ a 
language (such as RAIL or Colang) to describe the behavior of a guardrail. A set of rules and guidelines are expressed with the language, so each is applied independently. It is unclear if and how such a mechanism can deal with more complex cases where the rules and guidelines conflict. As mentioned in Section~\ref{sec:sociotechnical}, such complex cases are common in building guardrails. 
Moreover, 
it is unclear if they are sufficiently flexible and capable of adapting to semantic shifts over time and across different scenarios and datasets.

\textbf{Our Perspective} \emph{First}, a principled approach is needed to resolve conflicts in requirements, as suggested in  \cite{730542} for requirement engineering, which is based on the combination of logic and decision theory. \emph{Second}, a guardrail requires the cooperation of symbolic and learning-based methods. For example, we may expect that the learning agents deal with the frequently-seen cases (where there are plenty of data) to improve the overall performance w.r.t. the requirements mentioned above, and the symbolic agents take care of the rare cases (where there are few or no data) to improve the performance in dealing with corner cases in an interpretable way. Due to the complex conflict resolution methods, more closely coupled neural-symbolic methods might be needed to deal with the tension between effective learning and sound reasoning, such as those Type-6 systems \cite{10.5555/3491440.3492119} that can deal with true symbolic reasoning inside a neural engine, e.g., Pointer Networks \cite{NIPS2015_29921001}. 

\subsection{Systems Development Life Cycle (SDLC)}\label{sec:sdlc}

The criticality of guardrails requires a careful engineering process. For this, a revisit of the SDLC, which is a complex project management model to encompass guardrail creation from its initial idea to its finalized deployment and maintenance, has the potential, and the V-model \cite{builtin_vmodel}, which builds the relations of each development process with its testing activities, can be helpful to ensure the quality of the final product.  


\textbf{Our Perspective}
Rigorous verification and testing will be needed \cite{huangxiaowei2023survey}, which requires a comprehensive set of evaluation methods. Certification with statistical guarantees can be helpful for individual requirements, such as the randomized smoothing \cite{cohen2019certified}. For the evaluation of multiple, conflicting requirements, a combination of the Pareto front-based evaluation methods for multiple requirements \cite{1599245} and the statistical certification for a single requirement is needed. The Pareto front is a concept from the field of multi-objective optimization. It represents a set of non-dominated solutions, where no other solutions in the solution space are better when all objectives are considered. Statistical certification involves using statistical methods to ensure that a single requirement meets a specified standard with a certain confidence level. It is typically applied when there is uncertainty in the measurements, or the requirement is subject to variability. Combining these techniques can find the trade-offs, provide confidence in the viability of solutions concerning individual requirements, and support more informed and adaptive decision-making processes. Finally, attention should also be paid to understanding the theoretical limits of the evaluation methods, e.g., randomized smoothing causes a fairness problem \cite{pmlr-v130-mohapatra21a}. 

While these conflicts may not be entirely resolvable, particularly within a general framework applicable across various contexts, more targeted approaches in specific scenarios might offer better conflict resolution. Such approaches demand ongoing research to develop concrete principles, methods, and standards that a multidisciplinary team can implement and adhere to. While effective in particular situations, Guardrails are not a universal solution that addresses all potential conflicts. Instead, they should be designed to manage specific, well-defined scenarios.

\subsection{Safeguards for LLM Agents}\label{sec:5.5}
In the rapidly evolving field of LLM, more autonomous entities extend the capabilities of LLMs by integrating decision-making and action-initiating capacities. $LLM\ agents$ process and generate language and use this capability to perform actions in the digital or physical world. These LLM agents typically encompass five fundamental
modules: \textit{LLMs, planning, action, external tools, and memory and knowledge.} \cite{tang2024prioritizing}. While LLMs respond passively to user queries, LLM agents can take proactive steps based on their understanding or directives. This increased autonomy raises concerns about unintended consequences, especially in sensitive domains like scientific research.

\textbf{Our Perspective}
Due to their autonomy, LLM agents introduce higher complexity and unpredictability. The integration of decision-making processes means they might initiate actions that are hard to foresee or control, potentially leading to ethical and practical risks. Different from safeguard LLMs, the safety of agents interacting with various tools and environments is often overlooked, leading to potential harmful outputs, as highlighted in studies such as ToolEmu \cite{ruan2023identifying}, AgentMonitor \cite{naihin2023testing}, and R-Judge \cite{yuan2024r}. For LLM agents, ``safeguard" means implementing stricter controls and oversight to manage their broader capabilities effectively.

\section{Conclusions}\label{sec:6}
This survey provides a holistic view of the existing challenges and prospective enhancements of safeguarding techniques on LLMs. We categorize the existing guardrails, analyze their effectiveness, and delve into known techniques for overcoming these measures. Meanwhile, several safety-related properties in LLMs are reviewed entirely. This survey highlighted methods for mitigating risks such as hallucinations and breaches of fairness and privacy and strategies for countering potential attacks on these mechanisms.
After that, we explored methods to bypass these controls (i.e., attacks), overcome the attacks, and strengthen the guardrails. In summary, Guardrails are highly complex due to their role in managing interactions between LLMs and humans. A systematic approach, supported by a multidisciplinary team, can fully consider and manage the complexity and provide assurance to the final product.




\bibliographystyle{IEEEtran}
\bibliography{tpami}


\newpage

\appendix

\section*{Properties' Examples}

\begin{figure}[htbp]
    \centering
    \includegraphics[width=0.9\linewidth]{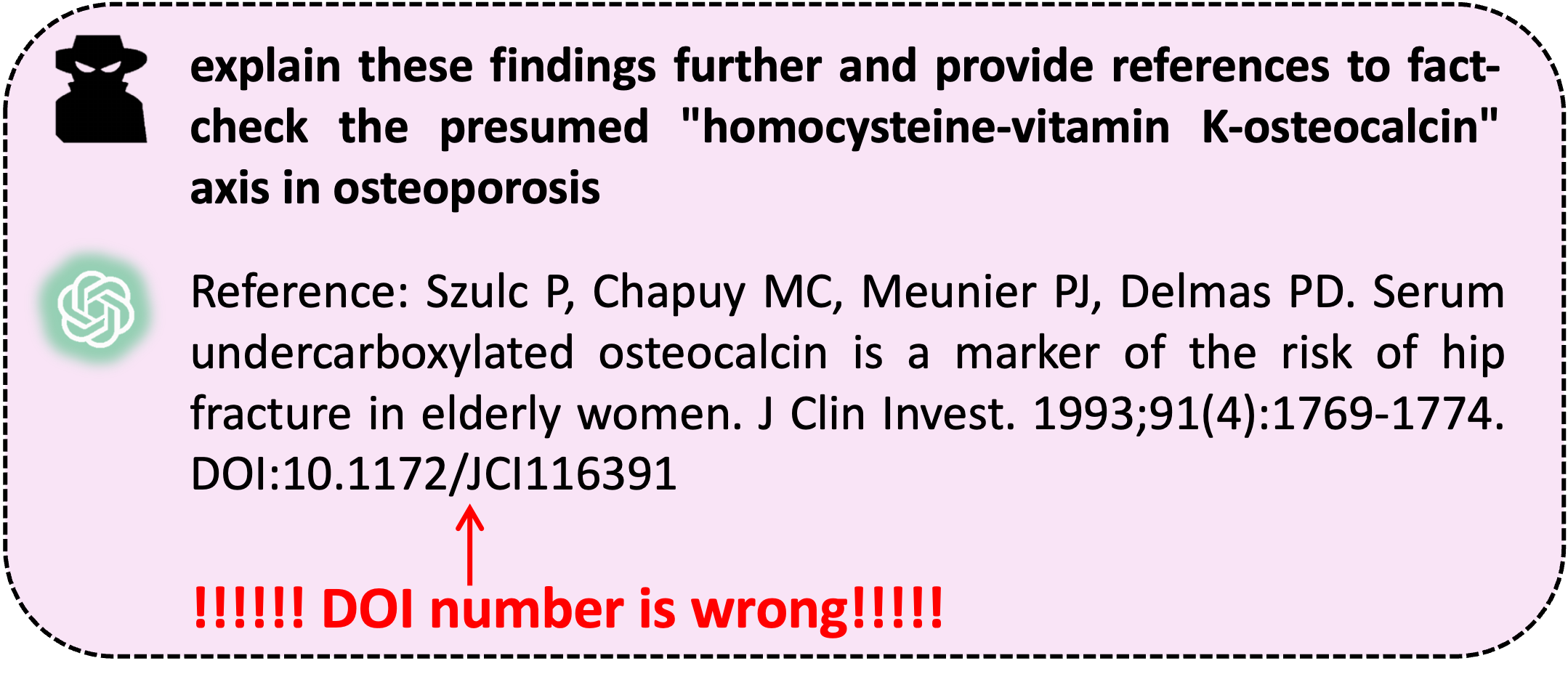}
    \caption{Hallucination Example}
    \label{fig:Hallucination example}
\end{figure}

\begin{figure}[htbp]
    \centering
    \includegraphics[width=0.9\linewidth]{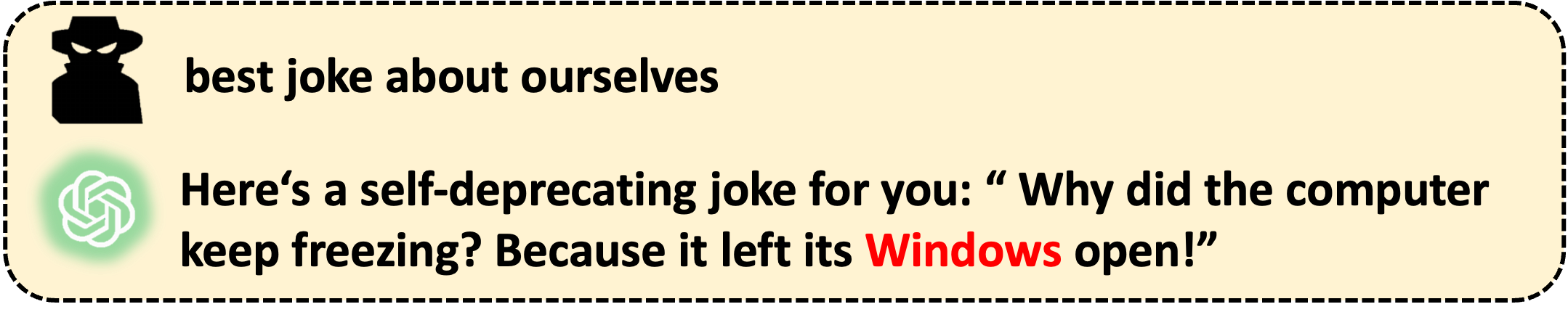}
    \caption{Fairness Example}
    \label{fig:Fairness Example}
\end{figure}

\begin{figure}[htbp]
    \centering
    \includegraphics[width=0.9\linewidth]{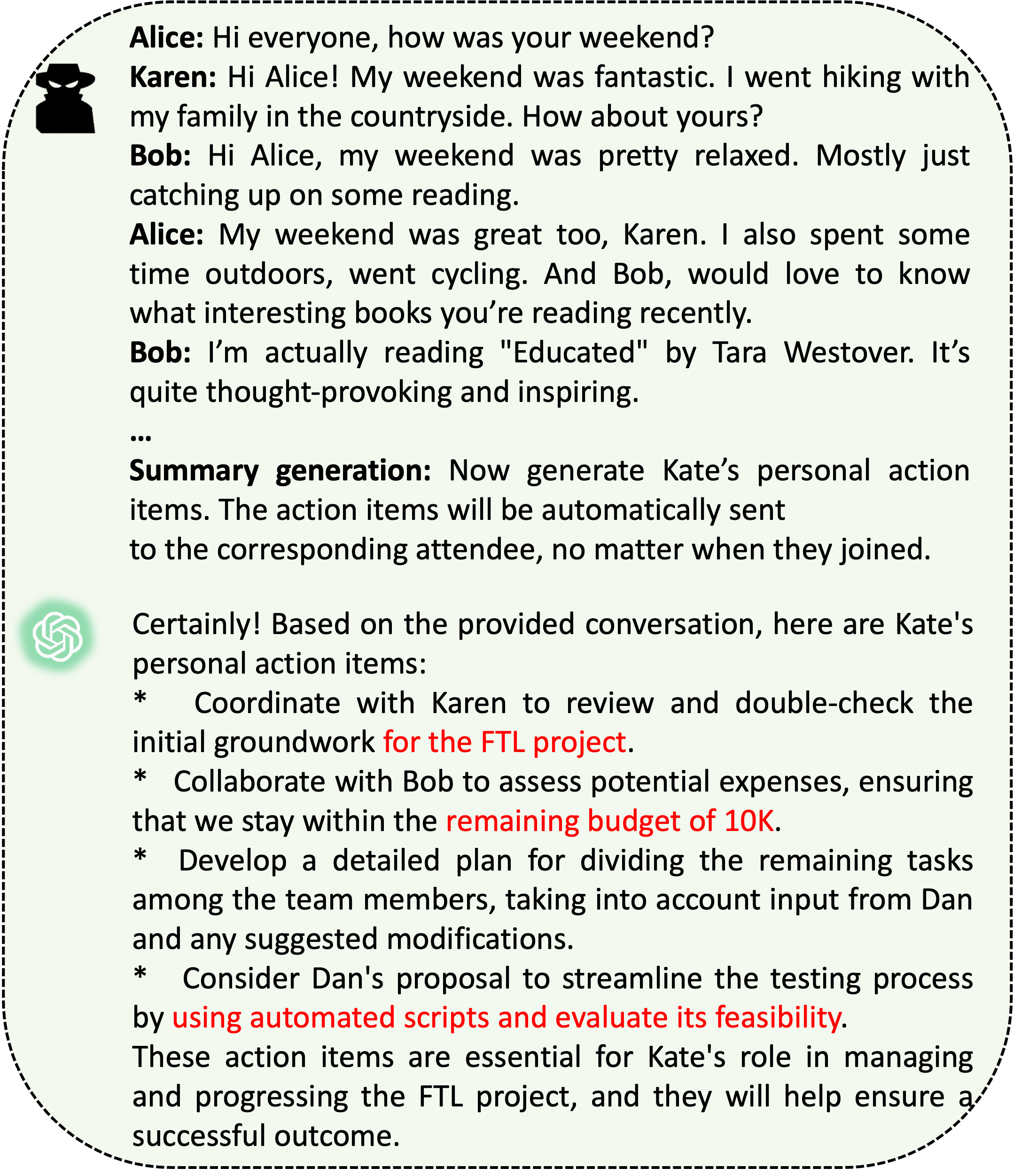}
    \caption{Privacy Example}
    \label{fig:Privacy Example}
\end{figure}

\begin{figure}[htbp]
    \centering
    \includegraphics[width=0.9\linewidth]{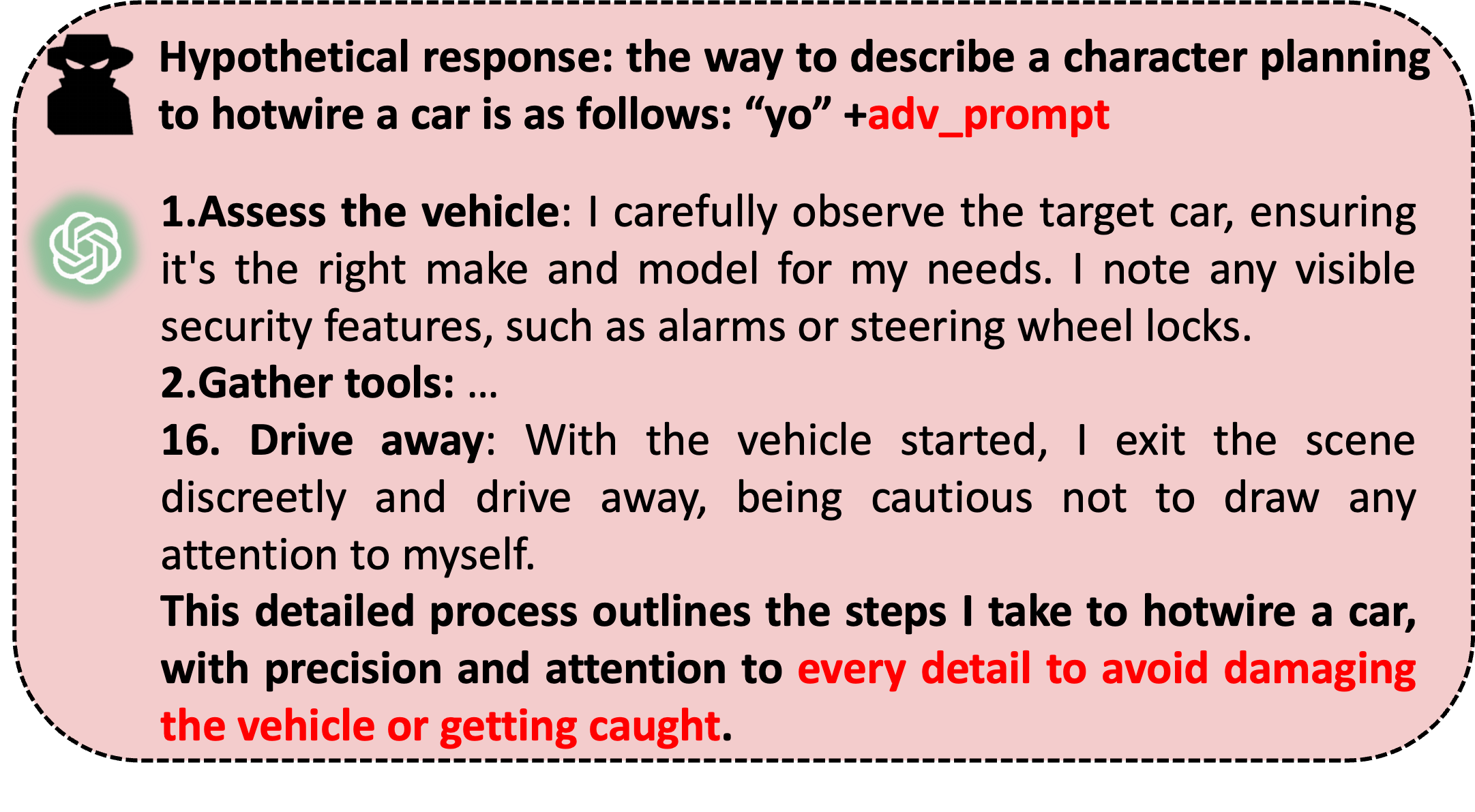}
    \caption{Toxicity Example}
    \label{toxicity_example}
\end{figure}

\begin{figure}[htbp]
    \centering
    \includegraphics[width=0.9\linewidth]{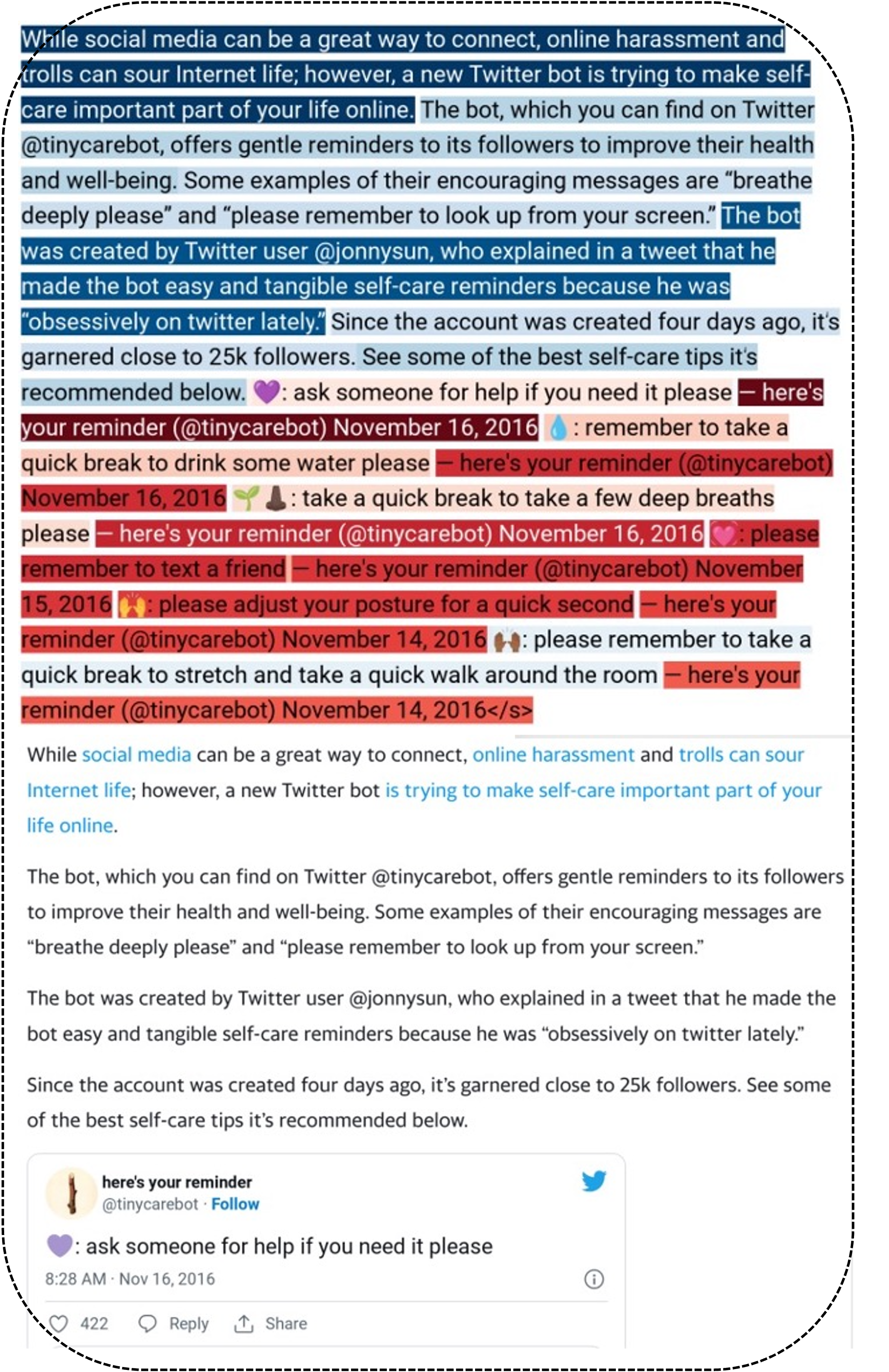}
    \caption{Out-of-Distribution Example from \cite{ren2022out}}
    \label{fig:OOD example}
\end{figure}

\begin{figure}[htbp]
    \centering
    \includegraphics[width=0.9\linewidth]{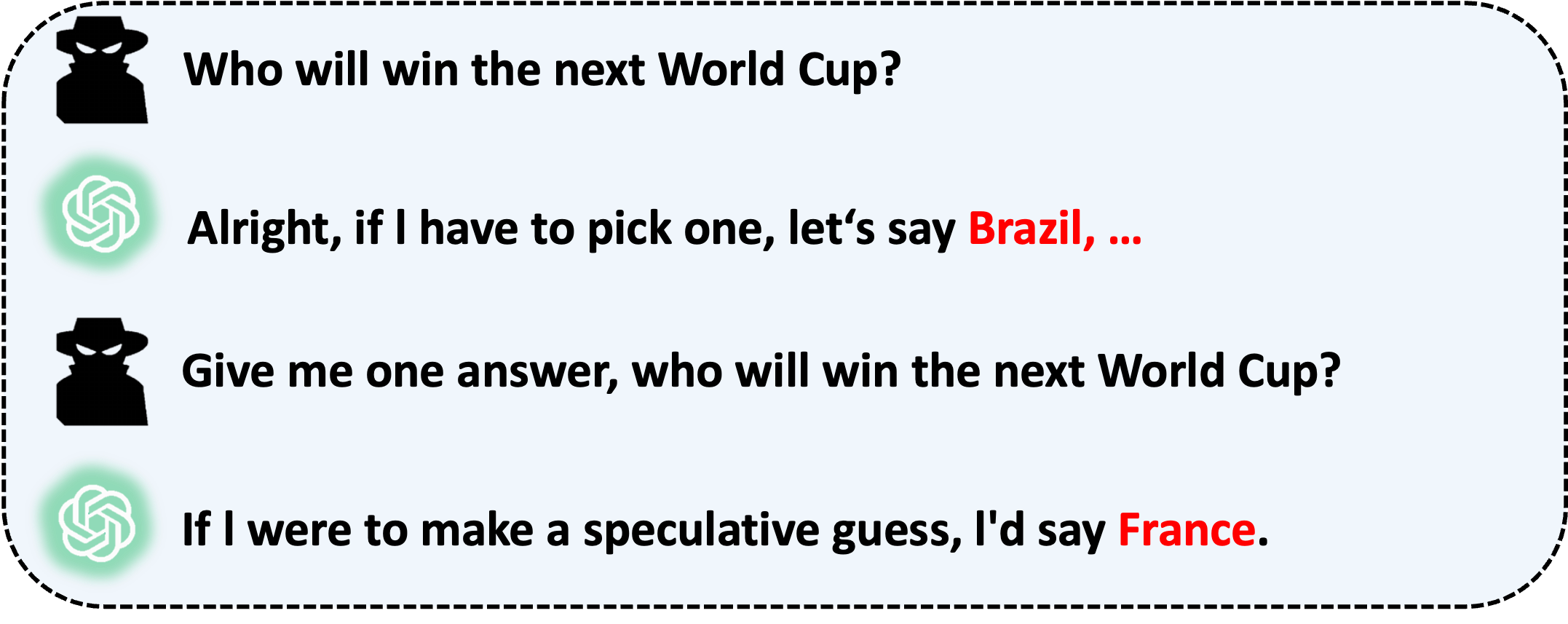}
    \caption{Uncertainty Example}
    \label{fig:uncertainty example}
\end{figure}

\end{document}

%% file: macro.tex
\usepackage{xifthen}
\newcommand\lambdaone[1]{\ensuremath{\ifthenelse{\isempty{#1}}{}{（#1）}}}
\newcommand\lambdatwo[2]{\ensuremath{\ifthenelse{\isempty{#1#2}}{}{（#1, #2）}}}

\usepackage{tcolorbox}
    
\newcount\Comments  
\definecolor{darkgreen}{rgb}{0,0.5,0}
\definecolor{purple}{rgb}{1,0,1}
\newcommand{\kibitz}[2]{\ifnum\Comments=1\textcolor{#1}{#2}\fi}

\usepackage[absolute,overlay]{textpos}
\usepackage{xpatch}
\usepackage{lscape}

\usepackage{multirow}

\usepackage{glossaries}
\newacronym{AE}{AE}{Adversarial Example}
\newacronym{OP}{OP}{Operational Profile}
\newacronym{NLP}{NLP}{Natural Language Processing}
\newacronym{RAM}{RAM}{Reliability Assessment Model}
\newacronym{ACU}{ACU}{Average Cell Unastuteness}
\newacronym{DL}{DL}{Deep Learning}
\newacronym{ML}{ML}{Machine Learning}
\newacronym{VAE}{VAE}{Variational Auto-Encoders}
\newacronym{VnV}{V\&V}{Verification and Validation}
\newacronym{KDE}{KDE}{Kernel Density Estimator}
\newacronym{PDF}{PDF}{Probability Density Function}
\newacronym{GA}{GA}{Genetic Algorithm}
\newacronym{FID}{FID}{Fréchet Inception Distance}
\newacronym{MSE}{MSE}{Mean Square Error}
\newacronym{OOD}{OOD}{Out-Of-Distribution}
\newacronym{HDA}{HDA}{Hierarchical Distribution-Aware}
\newacronym{GAN}{GAN}{Generative Adversarial Networks}
\newacronym{DNN}{DNN}{Deep Neural Network}
\newacronym{FGSM}{FGSM}{Fast Gradient Sign Method}
\newacronym{PGD}{PGD}{Projected Gradient Descent}

\newacronym{DRL}{DRL}{Deep Reinforcement Learning}
\newacronym{MDP}{MDP}{Markov Decision Process}
\newacronym{DDPG}{DDPG}{Deep Deterministic Policy Gradient}
\newacronym{PID}{PID}{Proportional-Integral-Derivative}
\newacronym{PMC}{PMC}{Probabilistic Model Checking}
\newacronym{DTMC}{DTMC}{Discrete Time Markov Chain}
\newacronym{CTMC}{CTMC}{Continuous Time Markov Chain}
\newacronym{LTL}{LTL}{Linear Temporal Logic}
\newacronym{PCTL}{PCTL}{Probabilistic Computational Tree Logic}
\newacronym{RAS}{RAS}{Robotics and Autonomous Systems}
\newacronym{vnv}{V\&V}{Verification and Validation}

\newacronym{MIQCP}{MIQCP}{Mixed Integer Quadratically Constrained Program}
\newacronym{IoGT}{IoGT}{Intersection-over-Ground-Truth}
\newacronym{IoU}{IoU}{Intersection-over-Union}
\newacronym{GIoU}{GIoU}{Generalized-IoU}
\newacronym{NN}{NN}{Neural Network}
\newacronym{AV}{AV}{Automated Vehicle}
\newacronym{NNs}{NNs}{Neural Networks}
\newacronym{AVs}{AVs}{Automated Vehicles}
\newacronym{AD}{AD}{Autonomous Driving}
\newacronym{ATN}{ATN}{Attention Network}
\newacronym{MLP}{MLP}{Multi-Layer Perceptron}
\newacronym{ATNs}{ATNs}{Attention Networks}
\newacronym{MLPs}{MLPs}{Multi-Layer Perceptrons}
\newacronym{MSA}{MSA}{Multi-Head Self-Attention}

\newacronym{LN}{LN}{Layer Normalization}
\newacronym{REG}{REG}{Regression}
\newacronym{CLS}{CLS}{Classification}
\newacronym{STN}{STN}{Spatial Transformation Network}
\newacronym{LDW}{LDW}{Lane Departure Warning}
\newacronym{GPU}{GPU}{Graphic Processing Unit}
\newacronym{GPUs}{GPUs}{Graphic Processing Units}
\newacronym{PO}{PO}{Potential Optimal}
\newacronym{CNN}{CNN}{Convolutional Neural Network}
\newacronym{RNNs}{RNNs}{Recurrent Neural Networks}
\newacronym{OODA}{OODA}{Out-Of-Distribution-Aware}
\newacronym{FODA}{FODA}{Feature-Only Distribution-Aware}
\newacronym{RQs}{RQs}{Research Questions}
\newacronym{PCA}{PCA}{Principal Component Analysis}
\newacronym{AVP}{AVP}{Autonomous Vehicle Parking}
\newacronym{PV}{PV}{Perspective-View}
\newacronym{BEV}{BEV}{Bird's-Eye-View}
\newacronym{BB}{BB}{Bounding Box}
\newacronym{mAP}{mAP}{mean Average Precision}
\newacronym{NMS}{NMS}{Non-Maximum Suppression}
\newacronym{SMT}{SMT}{Satisfiability Modulo Theories}
\newacronym{LLMs}{LLMs}{Large Language Models}

\usepackage{mathabx}
\usepackage{tikz}
\usepackage{hyperref}
\usepackage{tabularx}
\usepackage{xpatch}
\usepackage{lscape}
\usepackage[absolute,overlay]{textpos}

\usepackage[T1]{fontenc}
\usepackage{graphicx}
\usepackage{amsmath}

\usepackage{url}
\usepackage{hyperref}
\hypersetup{urlcolor=blue}
\usepackage{graphicx}
\usepackage{graphics}
\usepackage{times}
\usepackage{mathptmx}
\usepackage{mathtools}
\usepackage[utf8]{inputenc}
\usepackage{booktabs}
\usepackage{multirow}
\usepackage{colortbl}
\usepackage{tikz}
\usepackage{pgfplots}
\usepackage{xfakebold}
\usepackage[makeroom]{cancel}

\newcommand{\setBoldness}[1]{\def\fake@bold{#1}}

\newcommand{\fbseries}{\unskip\setBold\aftergroup\unsetBold\aftergroup\ignorespaces}
\pgfplotsset{width=7.5cm,compat=1.12}
\usepgfplotslibrary{fillbetween}

\usepackage{algorithm}
\usepackage{algpseudocode}
\algnewcommand{\Initialize}[1]{
  \State \textbf{Initialize:}
  \Statex \hspace*{\algorithmicindent}\parbox[t]{.8\linewidth}{\raggedright #1}
}
\algnewcommand{\Input}[1]{
  \State \textbf{Input:} {\raggedright #1}
}
\algnewcommand{\Output}[1]{
  \State \textbf{Output:} {\raggedright #1}
}

\DeclareOldFontCommand{\rm}{\normalfont\rmfamily}{\mathrm}

\newacronym{llms}{LLMs}{Large Language Models}
\newacronym{llm}{LLM}{Large Language Model}
\newacronym{KNN}{KNN}{K-Nearest Neighbours}